\documentclass[journal,twoside]{IEEEtran}
\usepackage[caption=false,font=normalsize,labelfont=sf,textfont=sf]{subfig}
\usepackage{mathrsfs}
\usepackage{latexsym}
\usepackage{graphicx}
\usepackage{epsfig}
\usepackage{subfig}
\usepackage{array}
\usepackage{amsmath}
\usepackage{amssymb}
\usepackage{color, soul}
\usepackage{amsthm}
\usepackage{enumerate}
\usepackage{caption}
\usepackage{algorithm}
\usepackage{algorithmic}
\renewcommand{\algorithmicrequire}{ \textbf{Inputs:}}
\renewcommand{\algorithmicensure}{ \textbf{Output:}}
\usepackage{url}

\usepackage{bm}
\usepackage{amssymb}
\usepackage{balance}
\usepackage{verbatim}
\usepackage{epstopdf}
\usepackage{setspace}
\usepackage{multicol}
\usepackage{amsmath,lipsum}
\usepackage{cuted}
\usepackage{multirow}
\usepackage{makecell}
\usepackage{verbatim}
\usepackage{graphicx}
\usepackage{color}
\usepackage{cite}
\usepackage {amssymb}
\usepackage{ragged2e} 
\usepackage{pifont}        
\usepackage[table]{xcolor} 
\usepackage{booktabs,makecell, multirow, tabularx}
\newcommand{\CheckmarkBold}{\textbf{\ding{51}}}  
\newcommand{\XSolidBold}{\textbf{\ding{55}}}     

\begin{document}
\title{Multi-user Downlink with Reconfigurable Intelligent Metasurface Antennas (RIMSA) Array

	\thanks{The authors are with the School of Information and Communications Engineering, Xi'an Jiaotong University,
		Xi'an 710049, China (e-mail:  wxj\_xjtu@163.com, xjbswhm@gmail.com).
	}
	\author{ Xuejian Wei, and Hui-Ming Wang, \emph{Senior Member, IEEE}
	}
}
\maketitle



\maketitle

\begin{abstract}
Reconfigurable Intelligent Surfaces (RIS) is a  transformative technology with great potential in many applications in  wireless communications and realizing the Internet of Everything at sixth generation (6G). In this study, we propose a wireless system  where the RIS acts as an antenna, which we call Reconfigurable Intelligent Metasurface Antennas (RIMSA). In particular, the base station (BS) equipped with a RIMSA array performs downlink transmissions to multiple users, where each user has a single or multiple RIMSA/RF links, and we aim to solve the sum-rate maximization problem by jointly optimizing the digital processing matrix of the transceivers and the phase responses of  RIMSA array at both BS and users.
For the multi-user multiple-input single-output (MU-MISO) scenario, we develop an alternating optimization algorithm  to slove the problem, where a fractional programming (FP) is used to optimize the digital processing matrix and a product manifold optimization (PMO) is proposed to provide the optimal phase responses of the RIMSA array at both BS and users.
For the multi-user multiple-input multiple-output (MU-MIMO) scenario,  we equate it to a weighted sum of mean square errors minimization problem, which can be solved by three subproblems iteratively. Both the optimal digital precoder subproblem and the optimal digital combiner subproblem have closed-form solutions, and the subproblem of RIMSA configuration is solved by
the PMO algorithm as well. Simulation results demonstrate that the proposed algorithms achieve significant performance gains over conventional
algorithms.
\end{abstract}


\section{Introduction}
In order to realize an intelligent information network for the Internet of Everything, the upcoming 6G wireless communication is expected to provide revolutionary mobile connectivity and high-throughput data communication services through low-power, highly integrated, and low-cost devices \cite{ref1}, \cite{ref2}. In order to realize the strong desire for ultra-high-speed data transmissions, large-scale multiple-input multiple-output (MIMO) technology and millimeter-wave (mmWave) communication have been widely studied and applied in recent years due to the fact that they can realize a higher data traffic through the wireless channel and provide a huge spectrum efficiency enhancement \cite{ref3}. Although MIMO technology is believed to  significantly promote network capacity by providing a large phased array, the inherent limitations of phased arrays prevent them from meeting this challenging vision of 6G networks. 
As the operating frequency of massive MIMO systems continues to increase, the hardware cost and power consumption of phased arrays are getting higher and higher, severely hindering their future application \cite{ref4}. As the number of antennas increases to an extra large massive MIMO (XL-MIMO), it leads to a excessive antenna size and limits the energy efficiency of the system \cite{refn8}. Therefore, there is an urgent need to develop more advanced antenna technologies to meet the future development of 6G and the exponentially increasing data demand in wireless communications \cite{ref6,ref7,ref8,ref9}.

In this context, the emerging Reconfigurable Intelligent Surfaces (RIS) \cite{ref12}  technology is considered as a promising energy  efficient communication assistant approach and has been widely studied in both academia and industry.
RIS is composed of a large number of low-cost passive reflecting elements, which realizes the intelligent regulation of parameters such as amplitude, phase, polarization, and frequency of electromagnetic waves \cite{refris}. They could be controlled adaptively to the wireless environment flexiblly. Generally, RIS can be categorized into reflective RIS, transmissive RIS and radiative RIS. 

Reflective RIS only passively reflects the incoming signals, eliminating the need for radio frequency (RF) transceiver hardware and complex signal processing \cite{refris1} to combat fading and enhance coverage . It is possible to artificially manipulate the phase and amplitude of the reflected signal so that the incident signal is reflected to the desired direction. In the last few years, there has been a great deal of work to demonstrate the superiority of RIS, such as capacity and communication rate improvement \cite{ref19,ref20,ref21}, energy efficiency optimization \cite{ref22}, and physical layer security enhancement \cite{ref24,ref25,refbai}, etc. Nevertheless, there is a double fading effect due to the reflective nature of this kind of RIS, which makes the received signals at the destination undergo a significant attenuation \cite{reflv}. To solve this problem, the authors in  \cite{reflv} and \cite{refD} proposed active RIS which can amplify the reflected signal. However, the extra power amplifiers increase the cost and power comsunption of the system. 

To increase the coverage area, the concept of transmissive RIS has received increasing attention \cite{refris1}. Instead of being reflected, the signal can be transmitted through the transmissive RIS to form a directional beam. In this way, transmissive RIS has the potential to fill the coverage gap of reflective RIS. However, the transmissive RIS has a very high transmission insertion loss \cite{Transris3}. As a result, the aperture efficiency of the transmissive RIS is significantly reduced.

Unlike the reflective and transmissive RIS, radiative RIS can regulate radiated or received waves and has signal processing capability, which can be deployed at the BS or user equipment (UE) side. Radiative RIS integrates both radiation and phase-shifting functions without the need for T/R components and phase shifters, which simplifies the cost and complexity of modulating radiated waves. 
Dynamic metasurface antennas (DMA) \cite{ref11} is one kind of radiative antenna. Architecturally, a DMA consists of a microstrip/waveguide containing multiple metamaterial elements arranged vertically, and one end of the microstrip is connected to the baseband signal processing unit via an RF chain. The singnals travel along waveguide and excite metamaterial elements successively, whose frequency response follows Lorentzian constraint.
Due to the propagation delay of the waveguide, it will exhibit some frequency selectivity. This will increase the complexity of the system  design and it is more difficult to control.
Reconfigurable holographic surfaces (RHS) \cite{ref7} is another type of radiative RIS antenna, which is a special leaky-wave antenna designed based on printed circuit board (PCB).
The electromagnetic waves (also known as reference waves) generated by the feed of the RHS propagates along the metasurface and radiates energy from the radiating elements. RHS utilizes metamaterials to construct holographic patterns on their surfaces based on the principle of holographic interference to produce the desired directional beam. The RHS mentioned above can only adjust the amplitude during transmission or reception of the signal.
Series feeds are employed in both DMA and RHS and the position of metamaterial
elements are further and further away from the feed point.

In addition, the study in \cite{FIM} introduces a novel type of radiating antenna termed the flexible intelligent metasurface (FIM), which is realized by depositing dielectric inclusions onto a conformal and flexible substrate. The position of each element can be dynamically adjusted in the direction normal to the surface through a process known as morphing. In contrast to conventional rigid metasurfaces, flexible intelligent metasurfaces (FIMs) offer a notable advantage in terms of mechanical adaptability, allowing seamless integration with non-planar or dynamically reconfigurable surfaces. This characteristic renders FIMs particularly suitable for deployment in structurally complex and time-varying environments where mechanical compliance is essential.
	Furthermore, to fully exploit the wireless propagation environment, stacked intelligent metasurface (SIM) \cite{SIM1} technology has been introduced to facilitate fine-grained EM manipulation. Unlike the aforementioned antennas, SIMs are composed of multiple layers of metasurfaces stacked together and strategically positioned in proximity to the transceivers. The multilayer architecture offers enhanced spatial gain and increased design flexibility, thereby enabling the synthesis of diverse  waveforms compared to their single-layer counterparts. This facilitates advanced functionalities such as MIMO precoding for channel capacity maximization \cite{SIM2}.
	However, the inherent multilayer structure of SIMs also introduces several implementation challenges. Notably, accurate calibration of inter-layer transmission coefficients and the design of wave-based beamforming (WBF) schemes require additional considerations compared to single-layer implementations.

\begin{figure}[!t]
	\centering
	\includegraphics[width=2.5in]{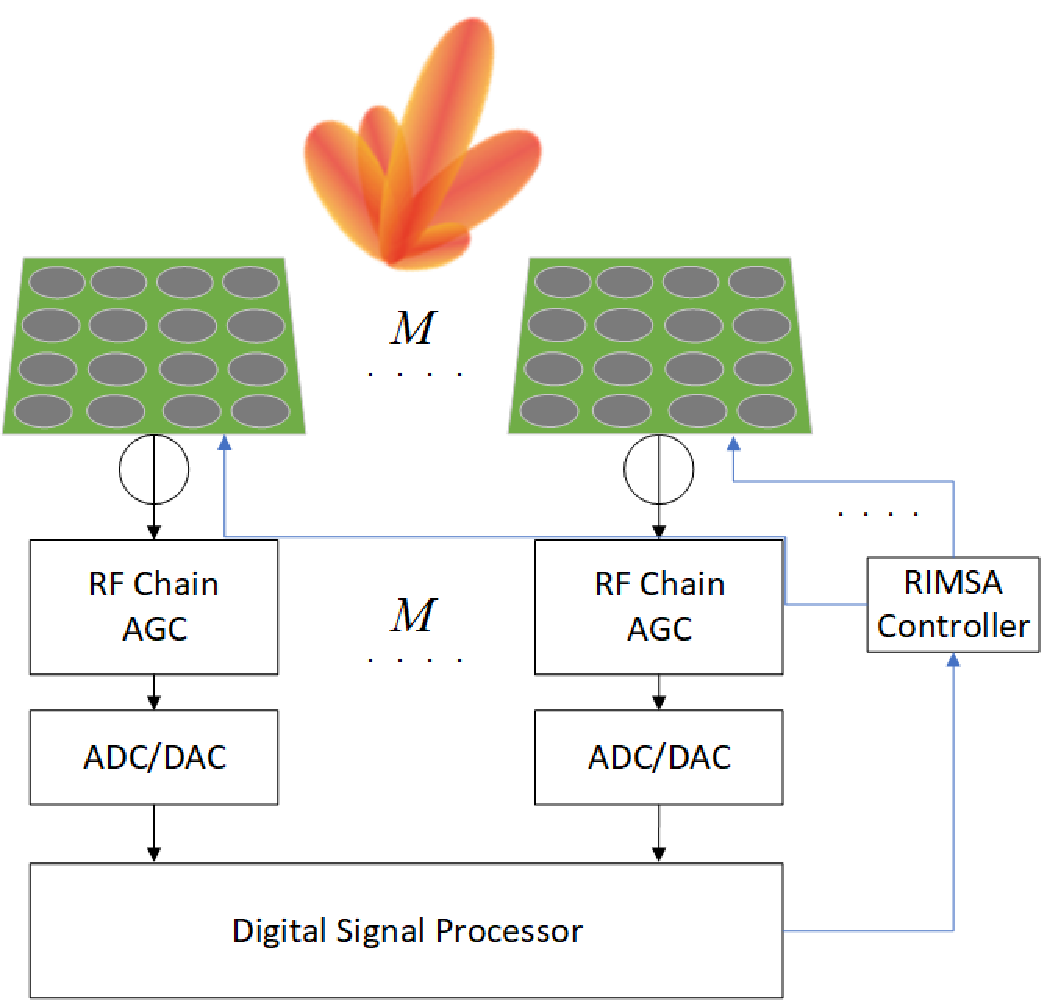}
	\caption{Transceiver structures based on RIMSA.}
	\label{fig_0}
\end{figure}

In this paper, a new architectural of radiative RIS is proposed, where all the metasurface elements are directly used as the radiating aperture of the antenna with a parallel coaxial feeding mode. This means all the metasurface elements will be excitated simultaneously via a power distribution network \cite{nat1}. This is significantly different from the radiative RIS mentioned above. We name this kind of antenna as Reconfigurable Intelligent Metasurface Antennas (RIMSA). 
The specific RMISA structure and the differences with the existing radiative RIS antennas are described in Section \ref{2}.
A wireless communication system based on RIMSA is investigated, as shown in Fig. \ref{fig_0}. 
Several metasurface elements connect an RF chain to the digital signal processor, which can control the phase of each metasurface element via an RIMSA controller. By combining the functionality of a programmable RIS element and a high-efficiency power radiator, a wide range of capabilities can be realized, including dynamic beam scanning capabilities. Such a structure avoids frequency selectivity and also has the capability of amplitude as well as phase modulation.

\subsection{Related Works and Motivations}
How to adjust the phase of the radiative RIS to maximize the capacity of the system is a very important issue.
For the research on DMA, the work in \cite{ref13} first develops a mathematical model of a BS equipped with the DMA as a transmit antenna under the downlink MU-MISO system. To maximize the sum rate, an efficient alternating optimization (AO) algorithm is proposed to optimizing the precoding matrix at the digital end and the DMA weights at the antenna end of the BS.
Later, the authors in \cite{ref14} investigates the weighted sum rate problem in downlink single-user multiple-input single-output (SU-MISO) and MU-MISO systems with DMA as the transmit antenna. An AO algorithm based on MMSE coding and manifold optimization (MO) algorithm is designed.
However the UEs are equipped with conventional antennas in these works.
Furthermore, the authors in \cite{ref15} studied the uplink MU-MIMO system where the DMA is used as a receiving antenna at the BS. It maximizes the energy efficiency of the system by jointly optimizing the transmit precoding matrix of the UE and the DMA weights of the BS based on the Dinkelbach transform and the AO method. 

 For RHS, the work in \cite{ref16} is the first to consider a model for multi-users communication with the RHS-equipped BS for the MU-MISO system.  The authors design a novel amplitude-controlled algorithm to maximize the sum rate, and a closed-form optimal holographic beamforming scheme is derived.
Unlike \cite{ref16}, UEs in \cite{ref17} are also equipped with the RHS.
Specifically, the authors proposed an AO algorithm that solves the digital beamforming subproblem at the BS, the holographic beamforming subproblem at the RHS, and the receive combining subproblem at the UE by using the power-allocated zero-forcing (ZF) method, the fractional programming (FP) method, and the coordinate ascent method \cite{ref17}, respectively. In addition, the work in \cite{ref18} extended the study to satellite communications. An uplink communication system consisting of one UE and multiple LEO satellites is considered and the communication rate maximization problem is constructed. The strict suboptimal ZF digital beamforming is adopted at the UE and holographic beamforming is optimized via programming. 

In \cite{FIM}, the FIM is employed as a transmitting antenna in a MISO system. Specifically, for the single-user scenario, each  element is configured to adjust its position in order to maximize the channel gain. The optimal surface profile is determined using an efficient Golden-Section Search (GSS) algorithm.
In contrast, for the multi-user scenario, the surface configuration at the base station must account for the signal-to-interference-plus-noise ratio (SINR) requirements of all users. A gradient descent algorithm is employed to optimize the FIM surface shape accordingly. This study considers a system where the FIM is deployed only at the transmitter side, while the receiver is equipped with a traditional antenna.

\subsection{Contributions and Paper Organization}
Based on the above discussion, it can be concluded that radiative RIS architectures proposed in prior studies suffer from inherent hardware limitations. Moreover, there is limited research focusing on sum-rate maximization for downlink communication in MU-MISO systems, and to the best of our knowledge, no existing work has addressed this problem in MU-MIMO downlink scenarios. 
Based on our new RIMSA, this paper focuses on the sum-rate maximization problem under downlink MU-MISO and MU-MIMO systems, where both the BS and UEs are equipped with the RIMSA. The main contributions of this paper are summarized as follows.
\begin{itemize}
\item[$\bullet$]
We proposes a novel radiative RIS, named RMSIA, and provides a detailed description of its structure along with a comparison of its advantages and disadvantages relative to existing radiative RIS.
 And we describe in detail the signal modeling of the RIMSA as transmitting and receiving antennas.
We are the first to propose MU-MISO and MU-MIMO models equipped with RIMSA at both the transmitter and receivers, and to present the sum rate maximization optimization problem, respectively.
\item[$\bullet$] Under the MU-MISO system, when optimizing the RIMSA phase shifts and digital precoders for transceivers, the deep coupling of the optimization variables due to multiuser rate summation and the high nonconvexity of the optimization variables under the constant mode constraints make the problem very complex and difficult to solve. 
To address this difficulty, we transform the original problem into a new one that has independent optimization variables. We then propose the product manifold optimization (PMO)  algorithm to simultaneously optimize the RIMSA phase shifts at the transceiver side.
In addition, we optimize the digital precoder using the FP algorithm and obtain the corresponding closed-form solutions.
\item[$\bullet$] In MU-MIMO systems, we transform the sum-rate maximization problem into a weighted sum MSE minimization problem due to the difficulty of direct optimization. The same problem as in MU-MISO systems exists in optimizing the RIMSA phase shifts for transceivers. To address these issues, we also transform the original problem into a new one where the optimization variables are independent in the new problem. Finally, we use the proposed PMO algorithm to simultaneously optimize the RIMSA configuration at the transceiver side.
In addition, a closed-form solution can be obtained by derivation when optimizing the digital precoder and combiner.
\end{itemize}

The rest of this paper is organized as follows. In Section \ref{2}, we present the system models for downlink MU-MISO and MU-MIMO in the case of applying RIMSA and the sum rate maximization problem, respectively.
Section \ref{3} proposes an AO algorithm to maximize the sum rate of MU-MISO system.
Section \ref{4} provides solutions under MU-MIMO system. 
In Section \ref{5}, we verify the performance of the proposed algorithm through simulation and compare it with existing baseline algorithms. Finally, Section \ref{6} summarizes the paper.

\subsection{Notations}
Throughout this paper, $a$ (or $A$), $\bf{a}$ and $\bf{A}$ stand for a
scaling factor, a column vector and a matrix, respectively. The transpose, conjugate transpose, and complex conjugate
are denoted by ${\left(  \cdot  \right)^T}$, ${\left(  \cdot  \right)^H}$ and ${\left(  \cdot  \right)^ * }$, respectively. 
$\left|  \cdot  \right|$, ${\left\|  \cdot  \right\|_F}$, ${\rm{Tr}}( \cdot )$ and ${( \cdot )^{ - 1}}$ denote the determinant (or module for a complex variable), Frobenius norm, trace and inverse of a matrix.
$ \odot $ is the Hadamard product of two matrices. 
${\mathbb{C}^{M \times N}}$ is a complex space with $M \times N$ dimensions.
$\mathbb{E}\left[  \cdot  \right]$ denotes
the expectation operator.
${{\bf{I}}_N}$ denotes the $N \times N$ identity matrix and ${{\bf{1}}_{N \times M}}$ denotes an $N \times M$ all-one matrix, respectively.
${\mathop{\rm Re}\nolimits} ( \cdot )$ represents the real part.
${\cal C}{\cal N}\left( {b,\sigma _{}^2} \right)$represents
the complex normal distribution with mean $b$ and variance ${\sigma _{}^2}$. ${\rm{blkdiag}}({{\bf{A}}_1},\cdots {{\bf{A}}_N})$ is a block diagonal matrix comprised of ${{\bf{A}}_1}, \cdots {{\bf{A}}_N}$.

\section{ System Model and Problem
Formulation}\label{2}
In this section, we present the signal model and problem formulation of the  downlink MU-MISO system and MU-MIMO system with RIMSA both at the BS and multiple UEs.

\subsection{RIMSA Architecture}
\begin{table*}
	[t!]
	\centering
	\label{tab1}
	\caption{ Comparison of RTPS and Traditional Phase Shifters}
	
	\resizebox{0.90\linewidth}{!}{
		\begin{tabular}{|c|c|c|}
			\hline
			Metric &Reflective-Type Phase Shifter (RTPS)  & Traditional Phase Shifters (T-type, $\pi$-type, Active)  \\\hline
			Control Method & PIN Diode / Varactor Diode& CMOS / MEMS / RF Amplifiers, etc.\\\hline
			Control Power Consumption &Extremely Low ($\mu$W level)& Relatively High (mW level or above)\\\hline
			Signal Path Power Loss & No Signal Power Loss&Noticeable Power Loss (especially for active types)\\\hline
			Suitable for Large Arrays & \CheckmarkBold & \XSolidBold\\\hline
		\end{tabular}
	}
	\end{table*}

We first give a brief introduction of our proposed RIMSA architecture.
Architecturally, the RIMSA is mainly consisted of a power distribution network, a phase control circuit, and a patch radiating antenna array. The power distribution network is responsible for distributing RF power signals of equal power to each RIMSA element. The phase control circuit is responsible for modulating the phase of the signal received by  or radiated from each RIMSA element. 
Specifically, each element incorporates a varactor diode and is connected to a branch of parallel feed network to directly radiate electromagnetic waves.
To enable integrated control of the radiating wave, the RIMSA provides a bias voltage channel for each metasurface element, allowing the phase response of each element to be controlled individually.
By controlling the direct current bias of the  phase modulation circuitry on each of the metasurface antenna elements, continuous phase control of the RF signal is achieved. Each RIMSA antenna is composed of multiple elements which is connected to a RF chain via coaxial cables. 

	RIMSA adopts the reflective-type phase shifter (RTPS). Compared with conventional transmission-type or active phase shifters, RTPS achieves significantly lower power consumption by avoiding active amplification or signal switching along the transmission path \cite{RTPS3}. The use of PIN or varactor diodes for phase tuning only requires minimal bias current or voltage, making RTPS highly suitable for energy-constrained and large-scale applications such as RIMSA arrays and UAV-assisted communication platforms. Several studies have provided theoretical and practical insights into the power consumption of RTPS, including models that consider the bias voltage and current in the case of PIN diodes \cite{RTPS2}, and the role of varactor diodes in reducing power dissipation \cite{RTPS1}. These findings highlight the low-power potential of RTPS, especially in systems with large numbers of elements, such as  RIMSA systems. 
	A comparison of the specifics is shown in the table I.
	
This architecture is similar to the partially connected hybrid architectures in massive MIMO \cite{HYB1,HYB2} but is realized via a number of metasurefaces. In addition, RIMSA consumes less energy, whereas the latter incurs additional hardware cost by using a unit-paradigm bounded phase shifter. This gives RIMSA a significant advantage in terms of hardware size and signal processing flexibility. 
And the architecture of RIMSA use bottom coaxial feed, where each metamaterial element is directly excited by the bottom feed. This architecture ensures better isolation between ports and makes it easier to realize 2D RIMSA than DMA. 
Furthermore, unlike RHS and DMA in which metamaterial elements are activated sequentially in the waveguide feed, RIMSA can excite all metamaterial elements simultaneously and through the coaxial feed network, which mitigates the frequency selectivity of the antenna response.

Compared to SIM, RIMSA offers a more practical and hardware-efficient solution for intelligent wireless transmission. While SIMs enable the construction of multi-layer metasurface architectures without direct connections between RF chains and metasurface elements \cite{sim,sim1}, thus providing enhanced spatial degrees of freedom and fine-grained electromagnetic wave control—they also introduce significant system complexity. This includes challenges in inter-layer electromagnetic coupling, intricate channel modeling, increased control overhead, and a higher demand for frequent CSI updates and advanced optimization algorithms, especially under dynamic wireless conditions. FIM manipulate EM by adjusting the positions of elements, which imposes stringent requirements on the mechanical structure of the antenna. In practice, achieving such precise control is challenging. In contrast, RIMSA adopts a single-layer structure with two-dimensional phase reconfiguration, striking a favorable balance between electromagnetic functionality and implementation complexity. This makes RIMSA a compelling alternative for realizing efficient, low-complexity intelligent antenna systems. In this paper, both the BS and UEs are equipped with RIMSAs.
\begin{figure}[!t]
	\centering
	\includegraphics[width=3.3in]{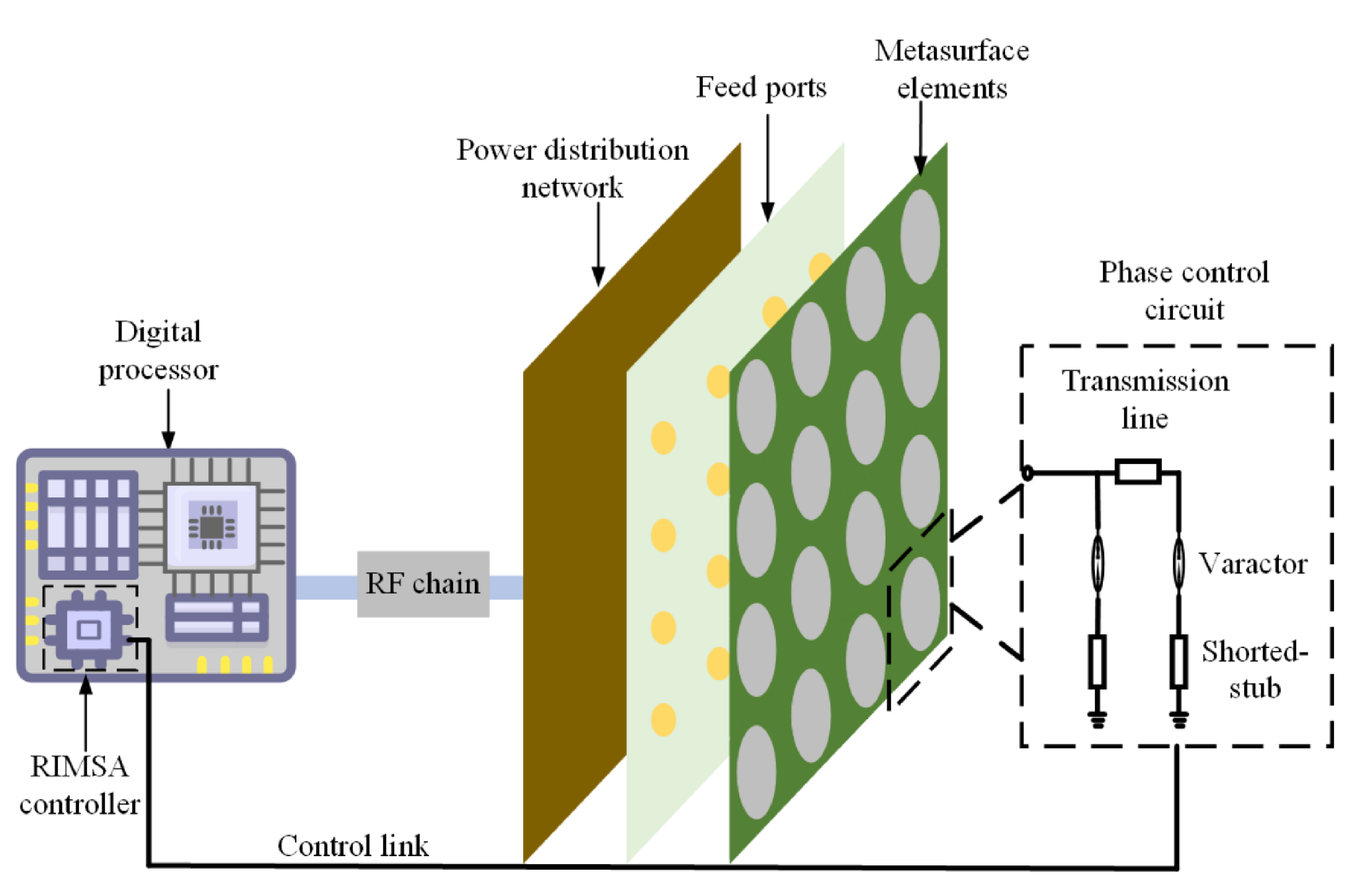}
	\caption{The architecture of RIMSA.}
	\label{fig_22}
\end{figure}

\subsection{System Model}
\subsubsection{Multi-User MISO System with RIMSA}
\begin{figure}[!t]
	\centering
	\includegraphics[width=3.4 in]{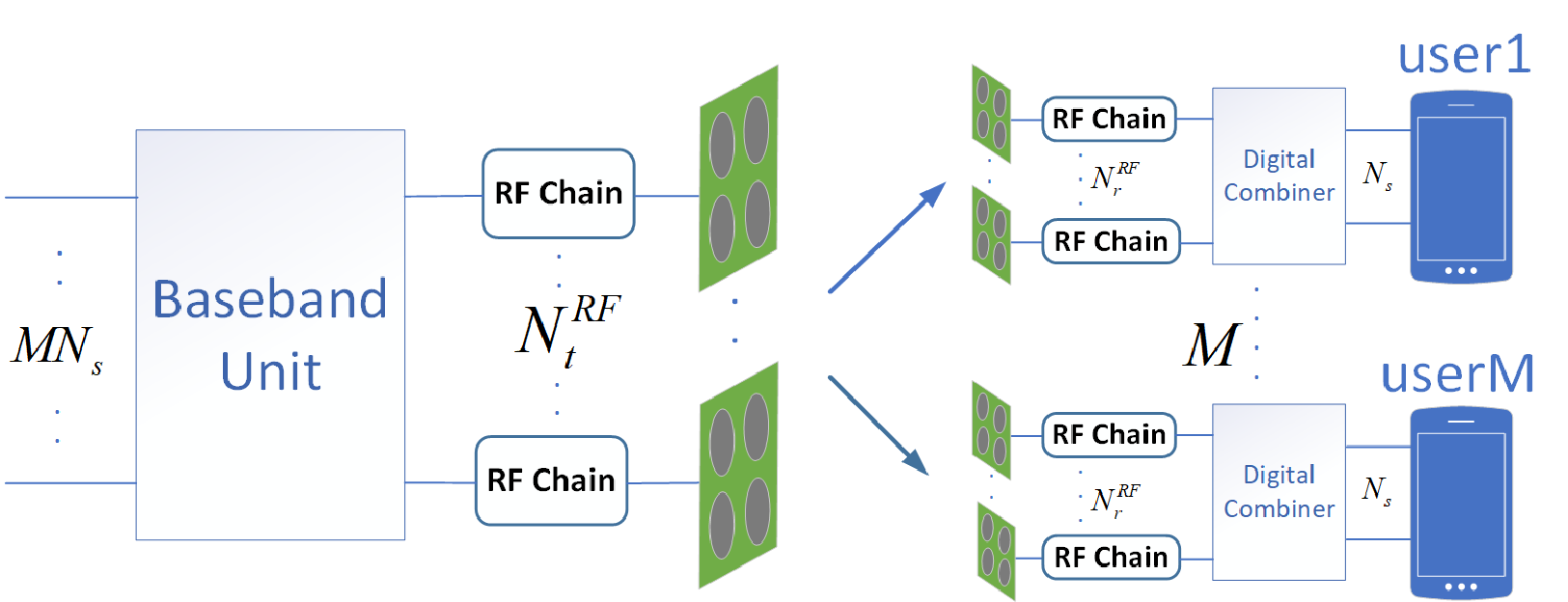}
	\caption{Downlink Multi-User MIMO System with RIMSA.}
	\label{fig_2}
\end{figure}

The specific MU-MISO system model is a special case of Fig. \ref{fig_2}. Each RIMSA is composed of ${N}$ metasurface elements, which as a whole is individually connected to a 
RF chain, and the BS 
is equipped with a RIMSA array with ${N_{RF}}$ RIMSA (and also ${N_{RF}}$ RF chains.).  Each RF chain is connected to the digital processor for signal processing via a AD/DA converter. The total number of RIMSA metasurface elements at the BS is ${N_t}=N_{RF} \cdot {N}$. There are ${M}$ UEs, and each UE is equipped with a RIMSA  consisting of ${N_r}$  metasurface elements. In other words,  the BS support ${M}$ data steams. 

For a downlink transmission, the baseband data stream ${\bf{s}} \in {\mathbb{C}^{M \times 1}}$ for all $M$-UEs first pass through the digital precoding matrix ${\bf{W}} = \left[ {{{\bf{w}}_1},{{\bf{w}}_2}, \cdots ,{{\bf{w}}_M}} \right] \in {\mathbb{C}^{{N_{RF}} \times M}}$, where ${{\bf{w}}_i} \in {\mathbb{C}^{{N_{RF}} \times 1}}, i = 1,2 \cdots M$ represents the precoding vector of the data stream to the ${i}$-th user. Then the signal is up-converted to the carrier frequency via the RF chain. After that, A  RIMSA  array is utilized to build the transmit beams. The beamformer at the RIMSA array ${\bf{V}} \in {\mathbb{C}^{N_t^{} \times {N_{RF}}}}$ is given by
\begin{equation}
\label{eq1}
{\rm{ }}{\bf{V}} = {\rm{blkdiag}}({{\bf{v}}_1},{{\bf{v}}_2} \cdots {{\bf{v}}_{{N_{RF}}}}),{\rm{  }}
\end{equation}
where ${{\bf{v}}_i} \in {\mathbb{C}^{N \times 1}}, {i = 1,2 \cdots {N_{RF}}}$. The element of ${{\bf{v}}_i}$ is constrained by the unit-modulus constraint as follows
\begin{equation}
\label{eq2}
{\cal V} = \left\{ {{{\bf{v}}_i}\left( j \right)\left| {\left| {{{\bf{v}}_i}(j)} \right| = 1,i = 1, \cdots, {N_{RF}},j = 1, \cdots, N} \right.} \right\},
\end{equation}
where ${{{\bf{v}}_i}\left( j \right)}$ is the  the ${j}$-th element of ${{\bf{v}}_i}$. Then the received signal of the ${m}$-th UE is
\begin{equation}
\label{eq3}
{y_m} = {\bf{f}}_m^H{{\bf{H}}_m}{\bf{V}}\sum\limits_{i = 1}^M {{{\bf{w}}_i}{s_i}}  + {\bf{f}}_m^H{\bf n}_m
\end{equation}
where ${{\bf{H}}_m} \in {\mathbb{C}^{{N_r} \times {N_t}}}$ represents the channel matrix between the RIMSA at the BS and the ${m}$-th UE, where each element of  ${{\bf{H}}_m}$ is the flat-fading channel coefficient between one pair of metasurface element,  and ${{\bf{f}}_m} \in {\mathbb{C}^{{N_r} \times 1}}$ represents the RIMSA phase response of the ${m}$-th UE. Similarly to (\ref{eq2}), each element in ${{\bf{f}}_i}$ is constrained by the unit-modulus constraint as follows
\begin{equation}
\label{eq5}
{\cal F} = \left\{ {{{\bf{f}}_i}\left( j \right)\left| {\left| {{{\bf{f}}_i}(j)} \right| = 1,i = 1, \cdots, M,j = 1, \cdots, {N_{r}}} \right.} \right\}.
\end{equation}
In (\ref{eq3}), ${{{s_i}}}$ denotes the transmit signal of the ${i}$-th UE, which is an independent random variable with
zero mean and unity variance. ${\bf n}_m$ is the additive Gaussian white noise vector at all the metafurface elements of the ${m}$-th UE with element variance ${\cal C}{\cal N}\left( {0,\sigma _m^2} \right)$.

The received signal interference noise ratio (SINR) of the ${m}$-th UE can be expressed as
\begin{equation}
\label{eq7}
{\rm{SIN}}{{\rm{R}}_m} = \frac{{{{\left| {{\bf{f}}_m^H{{\bf{H}}_m}{\bf{V}}{{\bf{w}}_m}} \right|}^2}}}{{\sum\limits_{i = 1,i \ne m}^M {{{\left| {{\bf{f}}_m^H{{\bf{H}}_m}{\bf{V}}{{\bf{w}}_i}} \right|}^2} +  \sigma _m^2{N_r}} }}.
\end{equation}
Since the response of each metasurface element satisfies the constant mode constraint, we get ${\bf{f}}_m^H{{\bf{f}}_m} = {N_r}$. 
Then the sum rate of the whole system can be written as $ {\rm{R = }}\sum\limits_{{{m = 1}}}^M {{\rm{lo}}{{\rm{g}}_{\rm{2}}}{\rm{(1 + SIN}}{{\rm{R}}_{{m}}}{\rm{)}}} $.

\subsubsection{Multi-User MIMO System with RIMSA}
For the MU-MIMO scenario, the BS is also equipped with a RIMSA array which has  $N_t^{RF}$ RF chains/RIMSAs, and each RIMSA is independently composed of $N$  metasurface elements. Then the total number of metasurface elements at the BS RIMSA is ${N_t} = N_t^{RF}  \cdot  N$. The BS serves $M$ UEs, and each UE is also similarly equipped with a RIMSA array with $N_r^{RF}$ RF chains/RIMSAs  to receive ${N_s}$ data streams. Each RIMSA is independently composed of $K$ metasurface elements, so the total number of elements at each UE is ${N_r}$ as ${N_r} = N_r^{RF} \cdot K$. the  system model is shown in Fig. \ref{fig_2}.

In this system, the BS sends precoded symbols with
digital precoder  ${\bf{W}}_i \in \mathbb{C}^{ N_t^{RF} \times {N_s}}, i = 1,2 \cdots M$ for the ${i}$-th UE and RIMSA array beamformer ${{\bf{V}}} \in {\mathbb{C}^{{N_t} \times N_t^{RF}}}$. Then the received signal of the $i$-th UE, $i = 1,2, \cdots M$ can be expressed as:
\begin{equation}
\label{eqm9}
\begin{split}
{{\bf{y}}_i} = &{{\bf{U}}_i}{\bf{W}}_{RF}^{\left( i \right)}{{\bf{H}}_i}{\bf{V}}{{\bf{W}}_i}{{\bf{x}}_i}+ \sum\limits_{\scriptstyle j = 1\hfill\atop
	\scriptstyle j \ne i\hfill}^M {{{\bf{U}}_i}{\bf{W}}_{RF}^{\left( i \right)}{{\bf{H}}_i}{\bf{V}}{{\bf{W}}_j}{{\bf{x}}_j}} \\& + {{\bf{U}}_i}{\bf{W}}_{RF}^{\left( i \right)}{{\bf{n}}_i},
\end{split}
\end{equation}
where  ${{\bf{x}}_i} \in {\mathbb{C}^{{N_s} \times 1}}$ represents the signal sent to the $i$-th user, ${{\bf{H}}_i} \in {\mathbb{C}^{{N_r} \times {N_t}}}$ represents the channel matrix between the BS and the ${i}$-th UE. And we assume $\mathbb{E}\left[ {{{\bf{x}}_i}{\bf{x}}_i^H} \right] = {{\bf{I}}_{{N_s}}}$. ${{\bf{n}}_i} \in {\mathbb{C}^{{N_s} \times 1}}$ is the additive Gaussian white noise at the $i$-th UE with ${\cal C}{\cal N}\left( {{\bf{0}},\sigma _i^2{{\bf{I}}_{{N_s}}}} \right)$. The RIMSA beamforming matrix ${\bf{V}}$ is the same as that of the MU-MISO system, which is also a block diagonal matrix as
\begin{equation}
\label{eqm10}
{\bf{V}} = {\rm{blkdiag}}({{\bf{v}}_1},{{\bf{v}}_2} \cdots {{\bf{v}}_{N_t^{RF}}}),
\end{equation}
where each element also satisfies the unit modulus constraint as in (\ref{eq2}). The range of {\bf{V}}  is denoted as ${\cal V}_M$. Similarly, the RIMSA receive beamforming matrix ${\bf{W}}_{RF}^{\left( i \right)} \in {\mathbb{C}^{N_r^{RF} \times {N_r}}}$ at the $i$-th UE is also a block diagonal matrix as:
\begin{equation}
\begin{split}
\label{eqm11}
&{\bf{W}}_{RF}^{\left( i \right)} =\\
&{\rm{blkdiag}}\left( {{{\left( {{\bf{w}}_{RF,1}^{\left( i \right)}} \right)}^H},{{\left( {{\bf{w}}_{RF,2}^{\left( i \right)}} \right)}^H} \cdots {{\left( {{\bf{w}}_{RF,N_r^{RF}}^{\left( i \right)}} \right)}^H}} \right),
\end{split}
\end{equation}
where ${\bf{w}}_{RF,j}^{\left( i \right)} \in {\mathbb{C}^{K \times 1}}$, $j = 1,2 \cdots N_r^{RF}$. The element of ${\bf{w}}_{RF,j}^{\left( i \right)}$ is also
constrained by the unit-modulus constrain. The range of ${\bf{W}}_{RF}^{\left( i \right)} $ is denoted as ${\cal W}_M$.
${{\bf{U}}_i} \in {\mathbb{C}^{N_s^{} \times N_r^{RF}}}$ represents the digital processing matrix of the $i$-th UE.

Then the rate of the $i$-th, $i = 1,2, \cdots M$ UE can be expressed as:
\begin{equation}
\label{eqm14}
\begin{split}
{r_i} = \log &\bigl| {{\bf{I}}_{{N_s}}} + {{\bf{U}}_i}{\bf{W}}_{RF}^{\left( i \right)}{{\bf{H}}_i}{\bf{V}}{{\bf{W}}_i} \times   \\
&{\bf{W}}_i^H{{\bf{V}}^H}{\bf{H}}_i^H{{\left( {{\bf{W}}_{RF}^{\left( i \right)}} \right)}^H}{\bf{U}}_i^H{\bf{R}}_{n,i}^{ - 1} \bigr|,
\end{split}
\end{equation}
where ${\bf{R}}_{n,i}=\sum\limits_{\scriptstyle j = 1\hfill\atop
\scriptstyle j \ne i\hfill}^M {{{\bf{U}}_i}{\bf{W}}_{RF}^{\left( i \right)}{{\bf{H}}_i}{\bf{V}}{{\bf{W}}_j}{\bf{W}}_j^H{{\bf{V}}^H}{\bf{H}}_i^H{{\left( {{\bf{W}}_{RF}^{\left( i \right)}} \right)}^H}{\bf{U}}_i^H}  \\+ \sigma _n^2{{\bf{U}}_i}{\bf{W}}_{RF}^i{\left( {{\bf{W}}_{RF}^{\left( i \right)}} \right)^H}{\bf{U}}_i^H$ represents the covariance matrix of the interference plus noise at the $i$-th UE, and the sum rate of the whole system can be writen as $
{{\rm{R}}_{\rm{M}}} = \sum\limits_{i{\rm{ = 1}}}^M {{r_i}} {\rm{   }}.$

\subsection{Channel Model}
In this paper, the propagation environment between each UE and the BS is modelled as a geometric narrowband downlink channel with ${L}$ paths \cite{ref27}.  We consider the  uniform planar array (UPA) model. This model depicts the channel matrix between the ${m}$-th user and the BS as:
\begin{equation}
\label{eq9}
{{\bf{H}}_m}{\rm{ = }}\sqrt {\frac{{{N_t}{N_r}}}{L}} \sum\limits_{\ell {\rm{ = 1}}}^L {\alpha _m^\ell } {{\bf{a}}_r}(\phi _{{r_m}}^\ell ,\theta _{{r_m}}^\ell ){{\bf{a}}_t}{(\phi _{{t_m}}^\ell ,\theta _{{t_m}}^\ell )^H},
\end{equation}
where $\alpha _m^\ell  \sim {\cal C}{\cal N}\left( {0,1} \right)$ is the complex gain of the ${\ell}$-th path between the BS and the user ${m}$. $\phi _{{r_m}}^\ell (\phi _{{t_m}}^\ell )$ and $\theta _{{r_m}}^\ell (\theta _{{t_m}}^\ell )$ stand for azimuth and elevation angles of arrival and departure (AoAs
and AoDs), respectively, and $\phi _{{r_m}}^\ell (\phi _{{t_m}}^\ell )\in\left[ {0,2\pi } \right)$, $\theta _{{r_m}}^\ell (\theta _{{t_m}}^\ell ) \in \left[ {0,2\pi } \right)$. In addition, ${{\bf{a}}_t}(.)$ and ${{\bf{a}}_r}(.)$ are the antenna array response vectors at the transmitter and receiver, respectively. In this paper, we consider the UPA with $N_y \times N_z =N$ intelligent metasurface elements and assuming that it is placed on the $y-z$ coordinate plane. Therefore, the array response vector can be written as:
{\begin{equation}
\begin{split}
\label{eq10}
{\bf{a}}(\phi ,\theta ) =& \sqrt {\frac{1}{N}} \bigl[ 1, \cdots ,{e^{jk\bar d\left( {p\sin \phi \sin \theta  + q\cos \theta } \right)}},\\ 
&\cdots ,{e^{jk\bar d\left( {( {N_y}  - 1)\sin \phi \sin \theta  + ( {N_z}  - 1)\cos \theta } \right)}}\bigl]^T
\end{split}
\end{equation} }
where $k = 2\pi /\lambda $, $\lambda$ is the wavelength of the transmitted signal and $\bar d$ represents the spacing between the intelligent surface elements, and $0 < p <  {N_y} $ and $0 < q <  {N_z} $ are the antenna indices in the 2D plane. This channel model will be used for simulation, but our tansceiver design and optimization hold for more general models.

\subsection{Problem Formulation}
\subsubsection{MU-MISO System}
In the MU-MISO system, our objective is to design the RIMSA beamforming matrices and the digital precoder at the BS, and the RIMSA receiving matrices at the UE to maximise the sum rate of the system. We assume that the perfect channel state information (CSI) is available \cite{ref28}. We can formulate the sum rate maximization problem as 
\begin{subequations}
\begin{align}
\mathop {\max }\limits_{{{\bf{V,W}},{{\bf{f}}_1},{{\bf{f}}_2} \cdots {{\bf{f}}_M}}} \quad&{\rm{     R}} = \sum\limits_{{\rm{m = 1}}}^M {{\rm{lo}}{{\rm{g}}_{\rm{2}}}{\rm{(1 + SIN}}{{\rm{R}}_m}{\rm{)}}}   \label{Za} \\
\mbox{s.t.}\quad
&{{\rm{  Tr}}({\bf{VW}}{{\bf{W}}^H}{\bf{V}}_{}^H) \le P},   \label{Zb}\\
&{\rm{            }}{\bf{V}} = {\rm{blkdiag}}({{\bf{v}}_1},{{\bf{v}}_2} \cdots {{\bf{v}}_{{N_{RF}}}}), \label{Zc} \\
&{\rm{            }}{{\bf{v}}_i}(j) \in {\cal V},{{\bf{f}}_i}(j) \in {\cal F},{\rm{ }}\forall i,\forall j  \label{Ze}
\end{align}
\end{subequations}
where ${P}$ is the transmit power. The diagonal matrix constraints (\ref{Zc}), and unit-modulus constraints (\ref{Ze}) are non-convex constraints, so this problem is in general difficult to tackle. 

\subsubsection{MU-MIMO System}
In the MU-MIMO system, our goal is to design the  RIMSA configurations matrices and baseband digital precoding matrices at both the BS and  $M$ UEs to maximize the sum rate of the system. We can formulate the sum rate problem as follows:
\begin{subequations}
	\begin{align}
		\mathop {\max }\limits_{{{\bf{V}}{\rm{, }}{\bf{W}}_D{\rm{,}}{\bf{W}}_{RF}^{\left( i \right)}{\rm{,}}{{\bf{U}}_i},\forall i}} \quad &{\rm{R}}_{\rm{M}} = \sum\limits_{i{\rm{ = 1}}}^M {{r_i}} \label{a}  \\
		\mbox{s.t.}\quad\quad
		&{{\rm{  Tr}}({\bf{V}}{{\bf{W}}}_D {{\bf{W}}_D^H}{\bf{V}}_{}^H) \le P},   \label{b} \\
		&{{\bf{V}}} \in {{\cal V}_M},{\bf{W}}_{RF}^{\left( i \right)} \in {{\cal W}_M},\label{c}
	\end{align}
\end{subequations}
where $P$ is the total transmit power and the digital precoding matrix ${\bf{W}}_D = \left[ {{{\bf{W}}_1},{{\bf{W}}_2}, \cdots ,{{\bf{W}}_M}} \right] \in {\mathbb{C}^{{N_t^{RF}} \times M{N_s}}}$. It can be seen that, like Problem (\ref{Za}), this problem has complex constraints, so it is a non-convex function, and the optimization variables are coupled in the constraints, resulting in a complex problem that is difficult to solve.

\section{Sum Rate Optimization for MU-MISO Systems}\label{3}
In this section, we propose an AO algorithm to solve problem (\ref{Za}). We will show that optimizing the digital precoding ${\bf{W}}$ in (\ref{Za}) is a fractional programming problem, which could be solved to get the optimal ${\bf{W}}$.
For the RIMSA configuration matrices ${{\bf{f}}_i},\forall i$ and ${\bf{V}}$, which are constrained on the manifolds, the PMO algorithm is proposed to optimize this nonconvex complex problem.

\subsection{Digital Baseband Precoder Design}
We first consider to design the digital precoder ${\bf{W}}$ when ${{\bf{f}}_m},\forall m$ and ${\bf{V}}$ are assumed to be fixed. Then ${\bf{\tilde h}}_m^H = {\left( {{{\bf{f}}_m^H}{{\bf{H}}_m}{\bf{V}}} \right)^H} \in {\mathbb{C}^{1 \times {N_{{\rm{RF}}}}}}$ can be considered as the equivalent channel from the BS to the ${m}$-th UE. Thus, the problem (\ref{Za}) for the digital precoder at the BS can be written as:
\begin{subequations}
\begin{align}
\label{eq12}
\mathop {\max }\limits_{\bf{W}}\quad &{\rm{R}} = \sum\limits_{m{\rm{ = 1}}}^M {{\rm{lo}}{{\rm{g}}_{\rm{2}}}\left( {{\rm{1 + }}\frac{{{{\left| {{\bf{\tilde h}}_m^H{{\bf{w}}_m}} \right|}^2}}}{{\sum\limits_{i = 1,i \ne m}^M {{{\left| {{\bf{\tilde h}}_m^H{{\bf{w}}_i}} \right|}^2} + \sigma _m^2{N_r}} }}} \right)} {\rm{, }}& \\
\mbox{s.t.}\quad
&{{\rm{  Tr}}({\bf{VW}}{{\bf{W}}^H}{\bf{V}}_{}^H) \le P}.
\end{align}
\end{subequations}
By introducing the auxiliary variable
 $\bm{\eta}  = [{\eta_1},{\eta_2}, \cdots {\eta _M}]$, 
we can show that the optimal solution of 
problem (\ref{eq12}) is equivalent to that of the following problem
\begin{subequations}
	\begin{align}
		\label{eq12a}
		&\mathop {\max }\limits_{{\bf{W}},{\bm{\eta }} } {\rm{   }} \sum\limits_{{{m = 1}}}^{{M}} {\left\{ {{\rm{lo}}{{\rm{g}}_{\rm{2}}}{\rm{(1 + }}{\eta _m}) - {\eta _m} + \left( {\frac{{{\rm{(1 + }}{\eta _m}){{\left| {{{{\bf{\tilde h}}}_m^H}{{\bf{w}}_m}} \right|}^2}}}{{\sum\limits_{i = 1}^M {{{\left| {{{{\bf{\tilde h}}}_m^H}{{\bf{w}}_i}} \right|}^2} + \sigma _m^2{N_r}} }}} \right)} \right\}} {\rm{,}} \\
		&\mbox{s.t.}\quad
		{{\rm{  Tr}}({\bf{VW}}{{\bf{W}}^H}{\bf{V}}_{}^H) \le P}.
	\end{align}
\end{subequations}
Indeed, the equivalence is due to the following fact:
For any given $\bf{W}$, (\ref{eq12a}) becomes a differentiable function
with respect to $\bm{\eta}$. Therefore, the optimal  $\bm{\eta}^o$ can be expressed as $\eta _m^o = \frac{{{{\left| {{\bf{\tilde h}}_m^H{{\bf{w}}_m}} \right|}^2}}}{{\sum\limits_{i = 1,i \ne m}^M {{{\left| {{\bf{\tilde h}}_m^H{{\bf{w}}_i}} \right|}^2} + \sigma _m^2{N_r}} }}$.
Substituting the optimal $\bm{\eta}^o$ into (\ref{eq12a}), it returns to the form
of problem (\ref{eq12}). 
Therefore, $\bf{W}$ obtained from problem (\ref{eq12a}) is the same as the
solution obtained from problem (\ref{eq12}).

Now we turn to solve (\ref{eq12a}). It is observed that for any given $\bm{\eta}$, the objective function is a fractional function with respect to ${{\bf{w}}_m}$, so it is a FP problem.
We can reformulate it as a biconvex optimisation problem as follows\cite{ref29}:
\begin{subequations}
	\begin{align}
		\label{eq13a}
 \mathop {\max }\limits_{{\bf{W}}{\rm{,}}{\bm{\alpha }}}\quad &{\rm{   }}{g_1} = \sum\limits_{m{\rm{ = 1}}}^M {2\sqrt {(1 + {\eta _m})} {\mathop{\rm Re}\nolimits} \left\{ {\alpha _m^*{\bf{\tilde h}}_m^H{{\bf{w}}_m}} \right\}}\\
 & {\rm{ + }}\sum\limits_{m{\rm{ = 1}}}^M {{{\left| {\alpha _m^{}} \right|}^2}\left( {\sum\limits_{i{\rm{ = 1}}}^M {{{\left| {{\bf{\tilde h}}_m^H{{\bf{w}}_i}} \right|}^2}}  + \sigma _m^2{N_r}} \right)} {\rm{,}} \notag\\
		\mbox{s.t.}\quad
		&{{\rm{  Tr}}({\bf{VW}}{{\bf{W}}^H}{\bf{V}}_{}^H) \le P},
	\end{align}
\end{subequations}
where ${\bm{\alpha }} = [{\alpha _1},{\alpha _2}, \cdots {\alpha _M}]$ is the auxiliary vector introduced
by the quadratic transformation. We can optimize $\bf{W}$ and ${\bm{\alpha }}$ alternatively.
To find $\bf{W}$, we define the Lagrangian function of this problem is
\begin{equation}
	\label{eql}
	{\cal L}\left( {{\bf{W}},{\bf{\alpha }},\lambda } \right){\rm{ = }}{g_1}{\rm{ -  }}\lambda  {\rm{Tr}}\left( {{\bf{VW}}{{\bf{W}}^H}{\bf{V}}_{}^H} \right){\rm{ + }}\lambda P,
\end{equation}
where $\lambda $ is the Lagrange multiplier. When $\bf{W}$ is fixed, by setting ${{\partial {\cal L}} \mathord{\left/
		{\vphantom {{\partial {\cal L}} {\partial {\alpha _m} = 0}}} \right.
		\kern-\nulldelimiterspace} {\partial {\alpha _m} = 0}}$, the auxiliary vector is
	\begin{equation}
		\label{eq222}
	{\alpha _m^o} = \frac{{(1 + {\eta _m}){\bf{\tilde h}}_m^H{{\bf{w}}_m}}}{{\sum\limits_{i{\rm{ = 1}}}^M {{{\left| {{\bf{\tilde h}}_m^H{{\bf{w}}_i}} \right|}^2}}  + \sigma _m^2{N_r}}},\forall m.
	\end{equation}
	Then, we fix ${\alpha _m^o}$ and apply Karush-Kuhn-Tucker (KKT) conditions to determine $\bf{W}$. By setting  ${{\partial {\cal L}} \mathord{\left/
		{\vphantom {{\partial {\cal L}} {\partial {{\bf{w}}_m} = 0}}} \right.
		\kern-\nulldelimiterspace} {\partial {{\bf{w}}_m} = 0}}$, we can obtain the optimal closed-form
solutions for ${{{\bf{w}}_m}}$ as
\begin{equation}
	\label{eq21}
{{\bf{w}}_m^o} = (1 + {\eta _m}){\alpha _m^o}{\left( {\sum\limits_{m{\rm{ = 1}}}^M {{{\left| {\alpha _m^{o}} \right|}^2}{\bf{\tilde h}}_m^{}{\bf{\tilde h}}_m^H + \lambda {\bf{V}}{{\bf{V}}^H}} } \right)^{ - 1}}{{\bf{\tilde h}}_m},
\end{equation}
In (\ref{eq21}), we determine the parameter $\lambda$ according to the transmitter power constraint :
\begin{equation}
	\label{eq211}
	\lambda  = \min \left\{ {\lambda  \ge 0:{\rm{Tr(}}{\bf{VW}}{{\bf{W}}^{\rm{H}}}{{\bf{V}}^{\rm{H}}}{\rm{)}} \le P} \right\}.
\end{equation}
It can be seen that ${\rm{Tr}}({\bf{VW}}{{\bf{W}}^{\rm{H}}}{\bf{V}}_{}^{\rm{H}})$ decreases monotonically as $\lambda$ increases. Therefore, the optimal $\lambda$ satisfying the power constraint is the minimum avaiable $\lambda$.
Specifically, when solving for ${\bf{W}}$ in practice, we initialise $\lambda$ and substitute it
into (\ref{eq21}), and then substitute ${\bf{W}}$
into (\ref{eq211}). If ${\rm{Tr}}({\bf{VW}}{{\bf{W}}^{\rm{H}}}{\bf{V}}_{}^{\rm{H}})$ is greater than $P$, we increase
$\lambda$; otherwise, we decrease $\lambda$. This process is repeated until a
specific stopping criterion is satisfied.
 Usually, we can obtain the optimal $\lambda$ by a bisection search \cite{ref30}.

\subsection{Optimization of RIMSA Phase Shifts for Transceivers }
In this subsection, we optimise both ${\bf{V}}$ and ${{\bf{f}}_m},\forall m$ with fixed ${\bf{W}}$. We propose a PMO algorithm based on the Riemannian gradient descent method to solve it.
\subsubsection{Problem Transformation} 
We can rewrite ${\rm{R }}$ in (\ref{Za}) as:
\begin{equation}
\label{eq24}
\begin{split}
{\rm{R }}&=\sum\limits_{m{\rm{ = 1}}}^M {{\rm{lo}}{{\rm{g}}_{\rm{2}}}{\rm{(1 + SIN}}{{\rm{R}}_m}{\rm{)}}}  \\
 &= \sum\limits_{m{\rm{ = 1}}}^M {{{\log }_2}\left( {\frac{{\sum\limits_{i = 1}^M {{{\left| {{{\bf{f}}_m^H}{{\bf{H}}_m}{\bf{V}}{{\bf{w}}_i}} \right|}^2} + \sigma _m^2{N_r}} }}{{\sum\limits_{i = 1,i \ne m}^M {{{\left| {{{\bf{f}}_m^H}{{\bf{H}}_m}{\bf{V}}{{\bf{w}}_i}} \right|}^2} + \sigma _m^2{N_r}} }}} \right)}  \\
& = {\log _2}\left( {\frac{{\det \left( {\left( {{\bf{\Phi }}{{\bf{\Phi }}^H}} \right) \odot {{\bf{I}}_M} + {{\bf{R}}_n}} \right)}}{{\det \left\{ {\left( {{\bf{\Phi }}\left( {{{\bf{\Phi }}^H} \odot {\bf{I}}_M^ - } \right)} \right) \odot {{\bf{I}}_M} + {{\bf{R}}_n}} \right\}}}} \right)\\
& =- {\log _2}\left( {\frac{{\det \left\{ {\left( {{\bf{\Phi }}\left( {{{\bf{\Phi }}^H} \odot {\bf{I}}_M^ - } \right)} \right) \odot {{\bf{I}}_M} + {{\bf{R}}_n}} \right\}}}{{\det \left( {\left( {{\bf{\Phi }}{{\bf{\Phi }}^H}} \right) \odot {{\bf{I}}_M} + {{\bf{R}}_n}} \right)}}} \right),
\end{split}
\end{equation}
where the received signal matrix ${\bf{\Phi }}$ we introduced as an intermediate variable, which is defined as follows:
\begin{equation}
\label{eq26}
\begin{split}
{\bf{\Phi }} &= {\bf{FHVW}} \\
&  = \left[ {\begin{array}{*{20}{c}}
{{\bf{f}}_1^H{{\bf{H}}_1}{\bf{V}}{{\bf{w}}_1}}&{{\bf{f}}_1^H{{\bf{H}}_1}{\bf{V}}{{\bf{w}}_2}}& \cdots &{{\bf{f}}_1^H{{\bf{H}}_1}{\bf{V}}{{\bf{w}}_M}}\\
{{\bf{f}}_2^H{{\bf{H}}_2}{\bf{V}}{{\bf{w}}_1}}&{{\bf{f}}_2^H{{\bf{H}}_2}{\bf{V}}{{\bf{w}}_2}}& \cdots &{{\bf{f}}_2^H{{\bf{H}}_2}{\bf{V}}{{\bf{w}}_M}}\\
 \vdots & \vdots & \ddots & \vdots \\
{{\bf{f}}_M^H{{\bf{H}}_M}{\bf{V}}{{\bf{w}}_1}}&{{\bf{f}}_M^H{{\bf{H}}_M}{\bf{V}}{{\bf{w}}_2}}& \cdots &{{\bf{f}}_M^H{{\bf{H}}_M}{\bf{V}}{{\bf{w}}_M}}
\end{array}} \right],
\end{split}
\end{equation}
where the $i$-th row of ${\bf{\Phi }}$ represents the signal received from the BS by the $i$-th UE,  ${i = 1,2 \cdots M}$. ${\bf{F}} \in {\mathbb{C}^{M \times M{N_r}}}$ is a block diagonal matrix  as $
{\bf{F}} = {\rm{blkdiag}}({\bf{f}}_1^H,{\bf{f}}_2^H \cdots {\bf{f}}_M^H)$,
and ${\bf{H}} = {\left[ {{\bf{H}}_1^T,{\bf{H}}_2^T, \cdots {\bf{H}}_M^T} \right]^T} \in {\mathbb{C}^{M{N_r} \times {N_t}}}$ represents the combined channel from the BS to all users.  ${{{\bf{I}}_M}}$ represents a unit matrix of dimension $M$. ${\bf{I}}_M^ -  \in {\mathbb{C}^{M \times M}}$ represents the matrix with all diagonal elements zero and all non-diagonal elements one. 
${{\bf{R}}_n} = {N_r} \cdot {\rm{blkdiag}}(\sigma _1^2,\sigma _2^2 \cdots \sigma _M^2) \in {\mathbb{C}^{M \times M}}$ represents the noise variance matrix of $M$ users. At this point, when ${\bf{W}}$ is fixed, maximizing the sum rate problem can be converted to:
\begin{subequations}
\begin{align}
\label{eq27}
\mathop {\min }\limits_{{\bf{V,F}}} \quad&{g_2}{\rm{(}}{\bf{V}},{\bf{F}}{\rm{) }} \\
\mbox{s.t.}\quad
&{\rm{            }}{\bf{V}} = {\rm{blkdiag}}({{\bf{v}}_1},{{\bf{v}}_2} \cdots {{\bf{v}}_{{N_{RF}}}}), \\
&{\rm{            }}{\bf{F}} = {\rm{blkdiag}}({{\bf{f}}_1},{{\bf{f}}_2} \cdots {{\bf{f}}_M}),{\rm{ }} \\
&{\rm{            }}{{\bf{v}}_i}(j) \in {\cal V},{{\bf{f}}_i}(j) \in {\cal F},{\rm{ }}\forall i,\forall j 
\end{align}
\end{subequations}
where the objective function ${g_2}$ can be expressed as:
\begin{equation}
\label{eq28}
\begin{split}
{g_2}\left( {{\bf{V}},{\bf{F}}} \right) &= {\log _2}\det \left\{ {\left( {{\bf{\Phi }}\left( {{{\bf{\Phi }}^H} \odot {\bf{I}}_M^ - } \right)} \right) \odot {{\bf{I}}_M} + {{\bf{R}}_n}} \right\}\\
&  - {\log _2}\det \left( {\left( {{\bf{\Phi }}{{\bf{\Phi }}^H}} \right) \odot {{\bf{I}}_M} + {{\bf{R}}_n}} \right).
\end{split}
\end{equation}

In this way, we have converted the  multiuser sum rate maximization problem into a  matrix multiplication minimization problem, eliminating the summation term and facilitating the subsequent derivation calculations.

\subsubsection{Proposed PMO Algorithm} 
The function ${g_2}$ is nonconvex and the constraints on the variables ${\bf{V}}$ and ${\bf{F}}$ are both nonconvex. It is difficult to solve this problem using traditional methods. In fact, define the sets in the constraints as ${{\cal S}_1} = \left\{ {{\bf{V}}:{\bf{V}} = {\rm{blkdiag}}({{\bf{v}}_1},{{\bf{v}}_2} \cdots {{\bf{v}}_{{N_{RF}}}}),{\rm{ }}\left| {{{\bf{v}}_i}(j)} \right| = 1,\forall i,j} \right\}$ and ${{\cal S}_2} = \left\{ {{\bf{F}}:{\bf{F}} = {\rm{blkdiag}}({{\bf{f}}_1},{{\bf{f}}_2} \cdots {{\bf{f}}_{{N_{RF}}}}),{\rm{ }}\left| {{{\bf{f}}_i}(j)} \right| = 1,\forall i,j} \right\}$, they are manifolds. And define the set ${\cal S} = {{\cal S}_1} \times {{\cal S}_2}$, which is known as the product of manifolds ${{\cal S}_1}$ and ${{\cal S}_2}$ \cite{ref31,ref32}.
We can see that the constraints for the variables ${\bf{V}}$ and ${\bf{F}}$ are on the manifold space   $\left( {{\bf{V}},{\bf{F}}} \right)\in {\cal S}$ and we can solve this complex problem based on the theory of manifold optimization.
In this paper, we propose the PMO algorithm based on the Riemann conjugate gradient algorithm to optimize ${\bf{V}}$ and ${\bf{F}}$ simultaneously.

The main idea of the PMO algorithm is to search for a sequence of $\left( {{{\bf{V}}_k},{{\bf{F}}_k}} \right),k = 0,1, \cdots  $ in the product manifold ${\cal S}$, such that the objective function ${g_2}{\rm{(}}{\bf{V}},{\bf{F}}{\rm{)}}$ is optimized with iterations. When convergence is reached, the PMO algorithm returns a solution that is the point at which the Riemannian gradient is zero. The solution is a suboptimal solution to problem (\ref{eq27}). The main advantage of this algorithm is that it can directly deal with non-convex (without any approximation) complex problems by means of the  theory of manifold optimization \cite{ref32} and it requires lower computational complexity compared to traditional methods.

We first give the tangent space of the manifold ${\cal S}$ at the point $\left( {{\bf{V}},{\bf{F}}} \right)$.  For any point ${\bf{V}} \in {{\cal S}_1}$ and ${\bf{F}} \in {{\cal S}_2}$, the tangent space passes tangentially through $\left( {{\bf{V}},{\bf{F}}} \right)$. Since manifold ${\cal S}$ is a product manifold, the corresponding tangent space ${T_{\left( {{\bf{V}},{\bf{F}}} \right)}}{\cal S}$ can be decomposed into the product of two tangent spaces, denoted by:
\begin{equation}
\label{eq29}
{T_{\left( {{\bf{V}},{\bf{F}}} \right)}}{\cal S} = {T_{\bf{V}}}{{\cal S}_1} \times {T_{\bf{F}}}{{\cal S}_2},
\end{equation}
where ${T_{\bf{V}}}{{\cal S}_1} = \left\{ {{{\bf{Z}}_1} \in {\mathbb{C}^{{N_t} \times {N_{RF}}}}\left| {{\mathop{\rm Re}\nolimits} \left\{ {{{\bf{Z}}_1} \odot {{\bf{V}}^*}} \right\} = {\bf{0}}} \right.} \right\},$ ${T_{\bf{F}}}{{\cal S}_2} = \left\{ {{{\bf{Z}}_2} \in {\mathbb{C}^{M \times M{N_r}}}\left| {{\mathop{\rm Re}\nolimits} \left\{ {{{\bf{Z}}_2} \odot {{\bf{F}}^*}} \right\} = {\bf{0}}} \right.} \right\},$ ${{\bf{Z}}_1}$ and ${{\bf{Z}}_2}$ are tangent vectors at points $\bf{V}$ and $\bf{F}$, respectively.

Under the concept of tangent space, the next key part of the PMO algorithm is to calculate the Riemannian gradient, which is the tangent direction in which the objective function decreases the fastest among all tangent vectors in the tangent space. Since ${{\cal S}_1} \in {\mathbb{C}^{{N_t} \times {N_{RF}}}}$ and ${{\cal S}_2} \in {\mathbb{C}^{M \times M{N_r}}}$ are subspaces of the Euclidean space, according to \cite{ref32}, the Riemannian  gradient can be computed as the orthogonal projection of the conventional Euclidean gradient into the Riemannian tangent space. Specifically,  it can be shown as:
\begin{equation}
\label{eq30}
gra{d_{{{\bf{V}}_k}}}{g_2} = Gra{d_{{{\bf{V}}_k}}}{g_2} - {\mathop{\rm Re}\nolimits} \left\{ {Gra{d_{{{\bf{V}}_k}}}{g_2} \odot {\bf{V}}_k^*} \right\} \odot {{\bf{V}}_k},
\end{equation}
\begin{equation}
\label{eq31}
gra{d_{{{\bf{F}}_k}}}{g_2} = Gra{d_{{{\bf{F}}_k}}}{g_2} - {\mathop{\rm Re}\nolimits} \left\{ {Gra{d_{{{\bf{F}}_k}}}{g_2} \odot {\bf{F}}_k^{\rm{*}}} \right\} \odot {{\bf{F}}_k},
\end{equation}
where $\left( {{{\bf{V}}_k},{{\bf{F}}_k}} \right)$ represents the iterative results obtained in the $k$-th search. $gra{d_{{{\bf{V}}_k}}}{g_2}$ stands for the Riemannian gradient of ${{\bf{V}}_k}$, and $Gra{d_{{{\bf{V}}_k}}}{g_2}$ is its Euclidean gradient. The same goes for ${gra{d_{{{\bf{F}}_k}}}{g_2}}$  and $ {Gra{d_{{{\bf{F}}_k}}}{g_2}}$.

Next we compute the Euclidean gradient of the objective function  $g_2$ with respect to  ${{\bf{F}}_k}$. The objective function is differentiated with respect to ${{\bf{F}}_k}$ and we can obtain the differential $d{g_2}$ as
\begin{equation}
\label{eq32}
{\rm{ }}d{g_2}{\rm{ = Tr}}\left\{ {{{\bf{A}}_k}\left( {d{{\bf{F}}_k}} \right)} \right\},
\end{equation} 
where  ${{\bf{A}}_k}$ is shown in equation (\ref{eq33}), which is calculated specifically in Appendix A. ${{\bf{F}}_k}$ is a block diagonal matrix and the Euclidean gradient can be expressed as:
\begin{equation}
\label{eq35}
Gra{d_{{{\bf{F}}_k}}}{g_2} = {\bf{A}}_k^H \odot {{\bf{P}}_1},
\end{equation}
where ${{\bf{P}}_1} = {\rm{blkdiag}}\left( {{{\left( {{\bf{p}}_1^{\left( 1 \right)}} \right)}^H},{{\left( {{\bf{p}}_1^{\left( 2 \right)}} \right)}^H} \cdots {{\left( {{\bf{p}}_1^{\left( M \right)}} \right)}^H}} \right)$ is a block diagonal matrix and has the same matrix structure as ${\bf{F}}$, and ${\bf{p}}_1^{\left( 1 \right)} = {\bf{p}}_1^{\left( 2 \right)} =  \cdots {\bf{p}}_1^{\left( M \right)} = {{\bf{1}}_{{N_r}}}$.

\begin{figure*}[ht] 
	\centering
	\begin{equation}
	\begin{split}
	{{\bf{A}}_k} &= {\bf{H}}{{\bf{V}}_k}{\bf{W}}\left( {{{\bf{\Phi }}^H} \odot {\bf{I}}_M^ - } \right)\left\{ {{{\left( {\left( {{\bf{\Phi }}\left( {{{\bf{\Phi }}^H} \odot {\bf{I}}_M^ - } \right)} \right) \odot {{\bf{I}}_M} + {{\bf{R}}_n}} \right)}^{ - 1}} \odot {{\bf{I}}_M}} \right\}
	- {\bf{H}}{{\bf{V}}_k}{\bf{W}}{{\bf{\Phi }}^H} \left\{ {{{\left( {\left( {{\bf{\Phi }}{{\bf{\Phi }}^H}} \right) \odot {{\bf{I}}_M} + {{\bf{R}}_n}} \right)}^{ - 1}} \odot {{\bf{I}}_M}} \right\}
	\label{eq33}
	\end{split}
	\end{equation}
	\begin{equation}
	\begin{split}
	{{\bf{B}}_k} &= {\bf{W}}\left( {{{\bf{\Phi }}^H} \odot {\bf{I}}_M^ - } \right)\left\{ {{{\left( {\left( {{\bf{\Phi }}\left( {{{\bf{\Phi }}^H} \odot {\bf{I}}_M^ - } \right)} \right) \odot {{\bf{I}}_M} + {{\bf{R}}_n}} \right)}^{ - 1}} \odot {{\bf{I}}_M}} \right\}{{\bf{F}}_k}{\bf{H}}
	- {\bf{W}}{{\bf{\Phi }}^H} \left\{ {{{\left( {\left( {{\bf{\Phi }}{{\bf{\Phi }}^H}} \right) \odot {{\bf{I}}_M} + {{\bf{R}}_n}} \right)}^{ - 1}} \odot {{\bf{I}}_M}} \right\}{{\bf{F}}_k}{\bf{H}}
	\label{eq34}
	\end{split}
	\end{equation}
	{\noindent} \rule[-10pt]{18cm}{0.05em}
\end{figure*}

Similarly,  the Euclidean gradient of the objective function ${g_2}$ to ${\bf{V}}$  can be expressed as:
\begin{equation}
	\label{eq42}
	 {Gra{d_{{{\bf{V}}_k}}}{g_2}} = {\bf{B}}_k^H \odot {{\bf{P}}_2},
\end{equation}
where ${{\bf{B}}_k}$ is shown in equation (\ref{eq34}), which is also calculated in the Appendix A.  ${{\bf{P}}_2} = {\rm{blkdiag}}({\bf{p}}_2^{\left( 1 \right)},{\bf{p}}_2^{\left( 2 \right)} \cdots {\bf{p}}_2^{\left( {{N_{RF}}} \right)})$ is a block diagonal matrix and has the same matrix structure as ${\bf{V}}$, and ${\bf{p}}_2^{\left( 1 \right)} = {\bf{p}}_2^{\left( 2 \right)} =  \cdots {\bf{p}}_2^{\left( {{N_{RF}}} \right)} = {{\bf{1}}_N}$.

After obtaining the Euclidean gradient $Gra{d_{{{\bf{F}}_k}}}{g_2} $ and $ Gra{d_{{{\bf{V}}_k}}}{g_2} $, the Riemannian gradient $gra{d_{{{\bf{F}}_k}}}{g_2}$ and $gra{d_{{{\bf{V}}_k}}}{g_2}$ can be expressed as the projection of the Euclidean gradient into the Riemannian domain, which can be calculated by (\ref{eq30}) and (\ref{eq31}).

With the Riemannian gradient, the next step is to find the search direction at the point $\left( {{{\bf{V}}_k},{{\bf{F}}_k}} \right)$. The search direction ${{\bf{\Pi }}_{{{\bf{F}}_k}}}$ and ${{\bf{\Pi }}_{{{\bf{V}}_k}}}$ at $\left( {{{\bf{F}}_k},{{\bf{V}}_k}} \right)$ should be updated at the iteration $k$  as
\begin{equation}
\begin{split}
\label{eq41}
{{\bf{\Pi }}_{{{\bf{F}}_k}}} = - gra{d_{{{\bf{F}}_k}}}{g_2} + \mu _k^{{\bf{F}}}{T_{{{\bf{F}}_{k - 1}} \to {{\bf{F}}_k}}}\left( {{{\bf{\Pi }}_{{{\bf{F}}_{k - 1}}}}} \right),
\end{split}
\end{equation}
\begin{equation}
	\begin{split}
		\label{eq43}
		{{\bf{\Pi }}_{{{\bf{V}}_k}}} = - gra{d_{{{\bf{V}}_k}}}{g_2} + \mu _k^{{\bf{V}}}{T_{{{\bf{V}}_{k - 1}} \to {{\bf{V}}_k}}}\left( {{{\bf{\Pi }}_{{{\bf{V}}_{k - 1}}}}} \right),
	\end{split}
\end{equation}
where ${T_{{{\bf{F}}_{k - 1}} \to {{\bf{F}}_k}}}\left( {{{\bf{\Pi }}_{{{\bf{F}}_{k - 1}}}}} \right)$ and ${T_{{{\bf{V}}_{k - 1}} \to {{\bf{V}}_k}}}\left( {{{\bf{\Pi }}_{{{\bf{V}}_{k - 1}}}}} \right)$ represents the  conversion of $ {{{\bf{\Pi }}_{{{\bf{F}}_{k - 1}}}}} $ on the manifold ${T_{{{{\bf{F}}_{k - 1}}} }}{\cal S}$ to the manifold ${T_{ {{{\bf{F}}_k}} }}{\cal S}$, and $ {{{\bf{\Pi }}_{{{\bf{V}}_{k - 1}}}}} $ on the manifold ${T_{{{{\bf{V}}_{k - 1}}} }}{\cal S}$ to the manifold ${T_{ {{{\bf{V}}_k}} }}{\cal S}$, respectively, which are defined in (\ref{eq45}). And $\mu _k^{{\bf{F}}}$ and $\mu _k^{{\bf{V}}}$ are Polak-Ribière parameter defined in (\ref{eq46}) \cite{ref32}, where ${\left\langle {.,.} \right\rangle _R}$ in (\ref{eq46}) is the Riemannian metric as
\begin{equation}
\label{eq47}
{\left\langle {{\bf{X}},{\bf{Y}}} \right\rangle _R} = {\mathop{\rm Re}\nolimits} \left( {{\rm{Tr}}\left( {{{\bf{X}}^H}{\bf{Y}}} \right)} \right).
\end{equation}
\begin{figure*}[ht] 
	\centering
	\begin{equation}
		\begin{split}
			{T_{{{\bf{Z}}_{k - 1}} \to {{\bf{Z}}_k}}}\left( {{{\bf{\Pi }}_{{{\bf{Z}}_{k - 1}}}}} \right) = {{\bf{\Pi }}_{{{\bf{Z}}_{k - 1}}}} - {\mathop{\rm Re}\nolimits} \left\{ {{{\bf{\Pi }}_{{{\bf{Z}}_{k - 1}}}} \odot {\bf{Z}}_k^*} \right\} \odot {\bf{Z}}_k^{},{\bf{Z}} \in \left\{ {{\bf{F}},{\bf{V}}} \right\}
			\label{eq45}
		\end{split}
	\end{equation}
	\begin{equation}
		\begin{split}
			\mu _k^{{\bf{Z}}} = \frac{{{{\left\langle {gra{d_{{{\bf{Z}}_k}}}{g_2},gra{d_{{{\bf{Z}}_k}}}{g_2} - {T_{{{\bf{Z}}_{k - 1}} \to {{\bf{Z}}_k}}}\left( {gra{d_{{{\bf{Z}}_{k - 1}}}}{g_2}} \right)} \right\rangle }_R}}}{{{{\left\langle {gra{d_{{{\bf{Z}}_{k - 1}}}}{g_2},gra{d_{{{\bf{Z}}_{k - 1}}}}{g_2}} \right\rangle }_R}}},{\bf{Z}} \in \left\{ {{\bf{F}},{\bf{V}}} \right\}
			\label{eq46}
		\end{split}
	\end{equation}
	{\noindent} \rule[-10pt]{18cm}{0.05em}
\end{figure*}
With the search direction ${{\bf{\Pi }}_{{{\bf{F}}_k}}}$ and ${{\bf{\Pi }}_{{{\bf{V}}_k}}}$, the iteration update is now to calculate ${{\bf{F}}_{k + 1}}$ and ${{\bf{V}}_{k + 1}}$ as
\begin{align}
{{\bf{F}}_{k + 1}} &={\cal R}\left(  {{\bf{F}}_k} + {\alpha _k}{{\bf{\Pi }}_{{{\bf{F}}_k}}}\right), \label{eq44}\\
{{\bf{V}}_{k + 1}} &={\cal R}\left(  {{\bf{V}}_k} + {\alpha _k}{{\bf{\Pi }}_{{{\bf{V}}_k}}}\right), 	\label{eq48}
\end{align}
where ${\alpha _k}$ is the step size of the $k$-th step of the search, which can be found using the backtracking line search algorithm \cite{ref32}. Note that the corresponding elements in the newly updated ${{\bf{F}}_{k + 1}}$  and ${{\bf{V}}_{k + 1}}$ may not satisfy the modulo one constraint, and therefore, a withdrawal operation is required during each iteration of the backtracking line search, defined as:
\begin{equation}
\label{eq49}
 {\cal R}({\bf{X}}(i,j)) = \frac{{{\bf{X}}(i,j)}}{{\left| {{\bf{X}}(i,j)} \right|}},\forall i,j.
  \end{equation}
This ensures that each non-zero element has a unit modulus.

\begin{algorithm}[t]
	\caption{PMO Algorithm for RIMSA Configuration}\label{alg:PMO}
	\begin{algorithmic}[t]
		\REQUIRE: $\varepsilon  > 0$, $c,\tau  \in \left( {0,1} \right)$ and initial point ${{\bf{F}}_0}$ and ${{\bf{V}}_0}$
		\STATE Set $k=0$, $\mu _0^{{\bf{F}}} =\mu _0^{{\bf{V}}}= 0$, ${{\bf{\Pi }}_{{{\bf{F}}_0}}} =  - gra{d_{{{\bf{F}}_0}}}{g_2}$ and ${{\bf{\Pi }}_{{{\bf{V}}_0}}} =  - gra{d_{{{\bf{V}}_0}}}{g_2}$
		\REPEAT
		\STATE Calculate search step size ${\alpha _k}$ according to backtracking line search algorithm, and find the next point ${{\bf{F}}_{k + 1}}$ and ${{\bf{V}}_{k + 1}}$  by (\ref{eq44}) and (\ref{eq48}).  
		\STATE Find the Riemannian gradient of the new point ${gra{d_{{{\bf{F}}_{k + 1}}}}{g_2}}$ and ${gra{d_{{{\bf{V}}_{k + 1}}}}{g_2}}$ according to Eq. (\ref{eq31}) and Eq. (\ref{eq30}).
		\STATE Calculate the vector transfer factor ${T_{{{\bf{F}}_k} \to {{\bf{F}}_{k + 1}}}}\left( {{{\bf{\Pi }}_{{{\bf{F}}_k}}}} \right)$ and ${T_{{{\bf{V}}_k} \to {{\bf{V}}_{k + 1}}}}\left( {{{\bf{\Pi }}_{{{\bf{V}}_k}}}} \right)$ through Eq. (\ref{eq45}).
		\STATE Calculate Polak-Ribière parameter $\mu _{k + 1}^{{\bf{F}}}$ and $\mu _{k + 1}^{{\bf{V}}}$ according to Eq. (\ref{eq46}).
		\STATE Calculate the search direction ${{\bf{\Pi }}_{{{\bf{F}}_{k + 1}}}}$ and ${{\bf{\Pi }}_{{{\bf{V}}_{k + 1}}}}$ by Eq. (\ref{eq41}) and Eq. (\ref{eq43}).
		\STATE $k=k+1$.
		\UNTIL{${\left\| {gra{d_{{{\bf{F}}_k}}}{g_2}} \right\|_F} + {\left\| {gra{d_{{{\bf{V}}_k}}}{g_2}} \right\|_F} < \varepsilon $}
	\end{algorithmic}
\end{algorithm}
\begin{algorithm}[t]
	\caption{ FP-PMO Algorithm for Multi-user MISO Systems }\label{alg2}
	\renewcommand{\algorithmicrequire}{\textbf{Input:}}
	\renewcommand{\algorithmicensure}{\textbf{Output:}}
	\begin{algorithmic}[1]
		\REQUIRE  ${\bf{H}}$ and ${{\bf{R}}_n}$. 
		\STATE Initialize ${{\bf{F}}_i},{{\bf{V}}_i},{{\bf{W}}_i}$, set $i=0$.
		\REPEAT
		\STATE Using the FP algorithm to calculate ${{\bf{W}}_i}$ by (\ref{eq21}). 
		\STATE Solving for ${{\bf{F}}_i}$ and ${{\bf{V}}_i}$ via Algorithm~\ref{alg:PMO}.
		\STATE $i=i+1$.
		\UNTIL{a stopping criterion triggers.}
		\ENSURE ${\bf{F}},{\bf{V}},{\bf{W}}$ 
	\end{algorithmic}
\end{algorithm}

We summarize the proposed PMO algorithm as Algorithm \ref{alg:PMO}. In this algorithm, $c$ is the percentage decrease in the acceptable value of ${g_2}$ in the backtracking line search, $\tau $ is the parameter used to regulate the change in the step size $\alpha $, and $\varepsilon $ is used to control the accuracy of the Riemannian gradient at convergence.
When the algorithm starts with $k = 0$, we first set the initial search direction ${{\bf{\Pi }}_{{{\bf{F}}_0}}} =  - gra{d_{{{\bf{F}}_0}}}{g_2}$ and ${{\bf{\Pi }}_{{{\bf{V}}_0}}} =  - gra{d_{{{\bf{V}}_0}}}{g_2}$,  and then compute the step size of this iteration by backtracking line search. If the Riemannian gradient reaches the target accuracy $\varepsilon $, the optimization algorithm stops, otherwise the above steps are repeated until the end condition is reached.

Up to now, we have proposed solutions for all optimized variables in problem (\ref{Za}). We now summarize the AO-PMO algorithm into Algorithm \ref{alg2}.

\subsection{Complexity Analysis}
As we can see, the main computational complexity of FP-PMO lies in computing (\ref{eq21}) and the  Euclidean gradient G$ra{d_{{{\bf{F}}_k}}}{g_2}$ and $Gra{d_{{{\bf{V}}_k}}}{g_2}$. Let ${I_{in}}$ and ${I_{out}}$ be the numbers of iterations of Algorithm~\ref{alg:PMO} and Algorithm~\ref{alg2}, respectively, and assume $M = {N_{RF}}$.
The complexity of (\ref{eq21}) is ${\cal O}\left( {{M^4}} \right)$ and the complexity of computing the  Euclidean gradient is ${\cal O}\left( {{M^2}{N_r}{N_t}} \right)$.
 So the complexity of the algorithm can be expressed as:
2${\cal O}\left( {{I_{in}}{I_{out}}{M^2}{N_r}{N_t}} \right)+{\cal O}\left( {{I_{out}}{M^4}} \right)$. 

\section{Sum Rate Optimization for MU-MIMO Systems}\label{4}
It can be seen that (\ref{a}) is a matrix determinant summation problem
and the deep coupling of the optimization variables due to multiuser rate summation and the high nonconvexity of the optimization variables under the constant mode constraints make the problem very complex and difficult to solve. 
In order to solve it, we first prove that the original optimization problem  can be equivalently converted to a WMMSE
optimization problem.
Then we fix ${\bf{V}}{\rm{,}}{\bf{W}}_{RF}^{\left( i \right)}$  and design the two digital processing matrices ${\bf{W}}_D$ and ${\bf{U}}_i,\forall i$
according to the WMMSE criterion, and they have closed-form
solutions. Subsequently, we fix the optimized ${\bf{W}}_D$ and ${\bf{U}}_i,\forall i$  and
jointly optimize ${\bf{V}}{\rm{,}}{\bf{W}}_{RF}^{\left( i \right)},\forall i$ by using
the PMO algorithm.

\subsection{Problem Transformation}
Our idea to solve  (\ref{a}) is to first transform
it into a WMMSE optimization problem. To this end, we 
define the MSE matrix for the $i$-th UE as
\begin{align}
		{\bf{E}}_i^{} &= \mathbb{E}\left[ {\left( {{{\bf{x}}_i} - {{\bf{y}}_i}} \right){{\left( {{{\bf{x}}_i} - {{\bf{y}}_i}} \right)}^H}} \right]\nonumber\\
		&= {{\bf{I}}_{{N_s}}} - {{\bf{U}}_i}{\bf{W}}_{RF}^{\left( i \right)}{{\bf{H}}_i}{\bf{V}}{{\bf{W}}_i} - {\bf{W}}_i^H{{\bf{V}}^H}{\bf{H}}_i^H{\left( {{\bf{W}}_{RF}^{\left( i \right)}} \right)^H}{\bf{U}}_i^H \nonumber\\
		& + \sum\limits_{j = 1}^M {{{\bf{U}}_i}{\bf{W}}_{RF}^{\left( i \right)}{{\bf{H}}_i}{\bf{V}}{{\bf{W}}_j}{\bf{W}}_j^H{{\bf{V}}^H}{\bf{{ H}}}_i^H{{\left( {{\bf{W}}_{RF}^{\left( i \right)}} \right)}^H}{\bf{U}}_i^H} \nonumber\\
		& + \sigma _n^2{{\bf{U}}_i}{\bf{W}}_{RF}^{\left( i \right)}{\left( {{\bf{W}}_{RF}^{\left( i \right)}} \right)^H}{\bf{U}}_i^H.\label{eq58}
\end{align}
Let ${{\bf{\Lambda }}_i}\succeq\bf{0}$ be the weight matrix of the $i$-th user. In Appendix B, we equivalently reformulate problem (\ref{a}) to the following weighted sum-MSE minimization problem as
\begin{subequations}
	\begin{align}
		\mathop {\max }\limits_{{{\bf{V}}{\rm{, }}{\bf{W}}_D{\rm{,}}{\bf{W}}_{RF}^{\left( i \right)}{\rm{,}}{{\bf{U}}_i},\forall i}} \quad &\sum\limits_{i{\rm{ = 1}}}^M {\left( {{\rm{Tr}}\left( {{{\bf{\Lambda }}_i}{{\bf{E}}_i}} \right) - \log \left| {{{\bf{\Lambda }}_i}} \right|} \right)}  \label{qa}  \\
		\mbox{s.t.}\quad\quad
		&{{\rm{  Tr}}({\bf{V}}{{\bf{W}}}_D {{\bf{W}}_D^H}{\bf{V}}_{}^H) \le P},   \label{qb} \\
		&{{\bf{V}}} \in {{\cal V}_M},{\bf{W}}_{RF}^{\left( i \right)} \in {{\cal W}_M}\label{qc}
	\end{align}
\end{subequations}
where the optimal solutions of ${\bf{V}}{\rm{,  }}{{\bf{W}}_{RF}^{\left( i \right)}}{\rm{,}}{\bf{W}}_D{\rm{,  }}{\bf{U}}_i$ in problem (\ref{qa})  and in problem (\ref{a})  are the same.  So we only need to solve the equivalence MSE minimization problem (\ref{qa}).
\subsection{Digital Precoding Design for Transceivers}

Firstly, it is easily observed that the objective function of (\ref{qa}) is convex with respect to each of the optimization variables ${\bf{W}}_i{\rm{,  }}{\bf{U}}_i$, and ${\bf{\Lambda }}_i, \forall i$, respectively. Therefore, they could be sloved by the alternation optimization (AO) principle.

Fixing the other four variables, ${{\bf{\Lambda }}_i}$ can be solved by using the block coordinate descent method, we have:
\begin{equation}
\label{eq63}
{\bf{\Lambda }}_i^{opt} = {\bf{E}}_i^{ - 1}.
\end{equation}

Fixing all optimization variables except ${\bf{U}}_i$, the well-known MMSE receiver can be derived  as
\begin{equation}
\label{eq60}
{\bf{U}}_i^{mmse} = {\left( {{\bf{J}}_i^{ - 1}{\bf{W}}_{RF}^{\left( i \right)}{{\bf{H}}_i}{\bf{V}}{{\bf{W}}_i}} \right)^H},
 \end{equation}
 where ${\bf{J}}_i^{} = \sum\limits_{j = 1}^M {{\bf{W}}_{RF}^{\left( i \right)}{{\bf{H}}_i}{\bf{V}}{{\bf{W}}_j}{\bf{W}}_j^H{{\bf{V}}^H}{\bf{H}}_i^H{{\left( {{\bf{W}}_{RF}^{\left( i \right)}} \right)}^H}}  + \sigma _n^2{\bf{W}}_{RF}^{\left( i \right)}{\left( {{\bf{W}}_{RF}^{\left( i \right)}} \right)^H}$ is the covariance matrix of the total received signal of the $i$-th UE. 

For ${\bf{W}}_i,\forall i$, fixing the other four variables, the objective function is a convex function with respect to ${\bf{W}}_i,\forall i$. The digital precoder can be derived by differentiating the objective function of (\ref{qa}) and set it to be zero.
The closed-form solution of $\left\{ {{{\bf{W}}_i}} \right\}_{i = 1}^M$  can be written as:
\begin{equation}
\begin{aligned}
\label{eq65}
&{\bf{W}}_i^{opt} = \\&{\left( {\sum\limits_{j = 1}^M {{{\bf{V}}^H}{\bf{H}}_{\rm{j}}^H{{\left( {{\bf{W}}_{RF}^{\left( j \right)}} \right)}^H}{\bf{U}}_j^H{{\bf{\Lambda }}_j}{\bf{U}}_j^{}{\bf{W}}_{RF}^{\left( j \right)}{{\bf{H}}_j}{\bf{V}}}  + {\mu _k}{{\bf{I}}_{N_t^{RF}}}} \right)^{ - 1}} \\
& \times {{\bf{V}}^H}{\bf{H}}_i^H{\left( {{\bf{W}}_{RF}^{\left( i \right)}} \right)^H}{\bf{U}}_i^H{{\bf{\Lambda }}_i}, 
\end{aligned}
\end{equation}
where ${\mu _k}$ should be chosen to satisfy the power constraint, just as in  (\ref{eq21}).
So this optimal value can be easily solved using one-dimensional search techniques (e.g., bisection method). Finally, by substituting ${\mu _k}$ into (\ref{eq65}), we obtain solutions for all ${\bf{W}}_i,\forall i$.

\begin{figure*}[t] 
	\centering

	\begin{equation}
	\begin{aligned}
	\label{eq68}
	f({\bf{V}}{\rm{,  }}{{\bf{W}}_{RF}^{\left( i \right)}}) &=\sum\limits_{i = 1}^M {{\rm{Tr}}\left[ {{{\bf{\Lambda }}_i}\left( {{{\bf{I}}_{{N_s}}} - {{\bf{U}}_i}{\bf{W}}_{RF}^{\left( i \right)}{{\bf{H}}_i}{\bf{V}}{{\bf{W}}_i} - {\bf{W}}_i^H{{\bf{V}}^H}{\bf{H}}_i^H{{\left( {{\bf{W}}_{RF}^{\left( i \right)}} \right)}^H}{\bf{U}}_i^H} \right)} \right]}\\&{\rm{ + }}\sum\limits_{i = 1}^M {{\rm{Tr}}\left[ {{{\bf{\Lambda }}_i}\left( {\sum\limits_{j = 1}^M {{{\bf{U}}_i}{\bf{W}}_{RF}^{\left( i \right)}{{\bf{H}}_i}{\bf{V}}{{\bf{W}}_j}{\bf{W}}_j^H{{\bf{V}}^H}{\bf{{ H}}}_i^H{{\left( {{\bf{W}}_{RF}^{\left( i \right)}} \right)}^H}{\bf{U}}_i^H} } \right)} \right]}{\rm{ + }}\sum\limits_{i = 1}^M {{\rm{Tr}}\left[ {K{{\bf{\Lambda }}_i}\sigma _i^2{{\bf{U}}_i}{\bf{U}}_i^H} \right]}.
	\end{aligned}
	\end{equation}
	\centering
	\begin{equation}
	\begin{aligned}
	\label{eq69}
	{f_1}({\bf{V}}{\rm{,  }}{{\bf{W}}_{RF}}) &= \sum\limits_{i = 1}^M {{\rm{Tr}}\left[ { - {{\bf{\Lambda }}_i}{{\bf{U}}_i}{\bf{W}}_{RF}^{\left( i \right)}{{\bf{H}}_i}{\bf{V}}{{\bf{W}}_i}} \right]} {\rm{ + }}\sum\limits_{i = 1}^M {{\rm{Tr}}\left[ {{{\bf{\Lambda }}_i}\left( {\sum\limits_{j = 1}^M {{{\bf{U}}_i}{\bf{W}}_{RF}^{\left( i \right)}{{\bf{H}}_i}{\bf{V}}{{\bf{W}}_j}{\bf{W}}_j^H{{\bf{V}}^H}{\bf{{ H}}}_i^H{{\left( {{\bf{W}}_{RF}^{\left( i \right)}} \right)}^H}{\bf{U}}_i^H} } \right)} \right]} \\
	&= - {\rm{Tr}}\left( {{\bf{\Lambda UW}}_{RF}^{}{\bf{HVW}}_D} \right) + {\rm{Tr}}\left( {{\bf{\Lambda UW}}_{RF}^{}{\bf{HVW}}_D{\bf{W}}_D^H{{\bf{V}}^H}{\bf{{ H}}}_{}^H{\bf{W}}_{RF}^H{{\bf{U}}^H}} \right).
	\end{aligned}
	\end{equation}
		{\noindent} \rule[-10pt]{18cm}{0.05em}
\end{figure*}

\subsection{Optimization of RIMSA Phase Shifts for Transceivers}
The digital precoder ${\bf{W}}_D$ and combiner ${\bf{U}}_i, \forall i$ have unconstrained solutions and can be derived in closed form. However, optimizing RIMSA phase shifts ${\bf{V}}$ and ${{\bf{W}}_{RF}^{\left( i \right)}}$ for transceivers with unit-modulus constraints is challenging. As was the case under the MU-MISO system, we will use the PMO algorithm to optimize ${\bf{V}}$ and ${{\bf{W}}_{RF}^{\left( i \right)}}$. With fixed ${\bf{W}}_D$, ${\bf{U}}_i$ and ${\bf{\Lambda }}_i$, the objective function of (\ref{qa}) can be transformed to (\ref{eq68}) where the second summation  holds due to ${\bf{W}}_{RF}^i{\left( {{\bf{W}}_{RF}^i} \right)^H} = K, \forall k$.

\subsubsection{Problem Transformation}
We  define a large matrix ${{\bf{W}}_{RF}} \in {\mathbb{C}^{MN_r^{RF} \times M{N_r}}}$ representing the RIMSA receive matrix for all UEs,  which is also a block diagonal matrix as
\begin{equation}
	\label{eqm12}
	{\bf{W}}_{RF}^{} = {\rm{blkdiag}}\left( {{\bf{W}}_{RF}^{\left( 1 \right)},{\bf{W}}_{RF}^{\left( 2 \right)} \cdots {\bf{W}}_{RF}^{\left(M \right)}} \right).
\end{equation}

Similarly, we define a large matrix ${\bf{U}} \in {\mathbb{C}^{^{MNs \times MN_r^{RF}}}}$ to include the digital processing matrices for all UEs. It can be represented as stacking each digital processing matrix ${{\bf{U}}_i}$  in a block diagonal manner as
\begin{equation}
	\label{eqm13}
	{\bf{U}} = {\rm{blkdiag}}({{\bf{U}}_1},{{\bf{U}}_2} \cdots {{\bf{U}}_M}).
\end{equation}
We also define ${\rm{ }}{\bf{\Lambda }} = {\rm{blkdiag}}({{\bf{\Lambda }}_1},{{\bf{\Lambda }}_2} \cdots {{\bf{\Lambda }}_M})$, and define ${\bf{H}} = {\left[ {{\bf{H}}_1^T,{\bf{H}}_2^T, \cdots {\bf{H}}_M^T} \right]^T} \in {\mathbb{C}^{M{N_r} \times {N_t}}}$ representing the combined channel from the BS to all users. 

After removing the terms in $f({\bf{V}}{\rm{,  }}{{\bf{W}}_{RF}})$ that are uncorrelated to ${\bf{V}}$ and ${{\bf{W}}_{RF}}$, we get objective function  ${f_1}({\bf{V}}{\rm{,  }}{{\bf{W}}_{RF}})$ in (\ref{eq69}). 
And the problem  can be transformed into:
\begin{equation}
	\begin{aligned}
		\label{eq67}
		\mathop {\min }\limits_{{\bf{V}}{\rm{,  }}{{\bf{W}}_{RF}}}\quad  &f_1({\bf{V}}{\rm{,  }}{{\bf{W}}_{RF}})     \\
		\mbox{s.t.}\quad
		&{\bf{W}}_{RF}^{} = {\rm{blkdiag}}\left( {{\bf{W}}_{RF}^{\left( 1 \right)},{\bf{W}}_{RF}^{\left( 2 \right)} \cdots {\bf{W}}_{RF}^{\left(M \right)}} \right),\\
		&{{\bf{V}}} \in {{\cal V}_M},{\bf{W}}_{RF}^{\left( i \right)} \in {{\cal W}_M},
	\end{aligned}
\end{equation}
We can see that the summation term is eliminated from the function ${f_1}({\bf{V}}{\rm{,  }}{{\bf{W}}_{RF}})$, which means that we can jointly optimize ${\bf{V}}$ and ${{\bf{W}}_{RF}}$ simultaneously. 

\subsubsection{Adopting PMO Algorithm} 
In the previous section, we have highlighted the theory of the PMO algorithm as well as the optimization process, so in this section we will not go into details. We only list the expressions for the Riemannian gradient of ${f_1}$ for ${\bf{V}}$ and ${{\bf{W}}_{RF}}$ as follows:
\begin{align}
	Gra{d_{{{\bf{V}}}}}{f_1} &= {\bf{C}}^H \odot {{\bf{Q}}_1}, 	\label{meq72}\\
	Gra{d_{{{{\bf{W}}_{RF}}}}}{f_1} &= {\bf{D}}^H \odot {{\bf{Q}}_2}, 	\label{meq73}
\end{align}
where ${\bf{C}} = {\bf{W}}_D{{\bf{W}}_D^H}{\bf{V}}_{}^H{\bf{{ H}}}_{}^H{\bf{W}}_{RF}^H{{\bf{U}}^H}{\bf{\Lambda UW}}_{RF}^{}{\bf{H}} - {\bf{W}}_D{\bf{\Lambda UW}}_{RF}^{}{\bf{H}}$.
 ${\bf{D}} = {\bf{HVW}}_D{{\bf{W}}_D^H}{\bf{V}}_{}^H{\bf{{ H}}}_{}^H{\bf{W}}_{RF}^H{{\bf{U}}^H}{\bf{\Lambda U}} - {\bf{HVW}}_D{\bf{\Lambda U}}$.
And ${{\bf{Q}}_1} = {\rm{blkdiag}}\left({\bf{q}}_1^{\left( 1 \right)},{\bf{q}}_1^{\left( 2 \right)} \cdots {\bf{q}}_1^{\left( {{N_t^{RF}}} \right)}\right)$ is a block diagonal matrix and has the same matrix structure as ${\bf{V}}$, and  ${\bf{q}}_1^{\left( 1 \right)} = {\bf{q}}_1^{\left( 2 \right)} =  \cdots {\bf{q}}_1^{\left( {N_t^{RF}} \right)} = {{\bf{1}}_N}$. 
 ${{\bf{Q}}_2} = {\rm{blkdiag}}\left( {{{\left( {{\bf{q}}_2^{\left( 1 \right)}} \right)}^H},{{\left( {{\bf{q}}_2^{\left( 2 \right)}} \right)}^H} \cdots {{\left( {{\bf{q}}_2^{\left( {MN_r^{RF}} \right)}} \right)}^H}} \right)$ is a block diagonal matrix with $MN_r^{RF}$ rows and has the same matrix structure as ${{\bf{W}}_{RF}}$, and ${\bf{q}}_2^{\left( 1 \right)} = {\bf{q}}_2^{\left( 2 \right)} =  \cdots {\bf{q}}_2^{\left( {N_t^{RF}} \right)} = {{\bf{1}}_K}$.
The WMMSE-PMO algorithm is summarized as Algorithm \ref{alg3}.

\subsection{Complexity Analysis}
As we can see, the main computational complexity of WMMSE-PMO lies on computing (\ref{eq63}), (\ref{eq60}), (\ref{eq65}) and the Euclidean gradient $Gra{d_{{{\bf{V}}}}}{f_1}$ and $Gra{d_{{{{\bf{W}}_{RF}}}}}{f_1}$, which are 
${\cal O}\left( {MN_t^{RF}{N_s}{N_t}} \right)$,
${\cal O}\left( {MN_t^{RF}{N_r^{RF}}{N_t}} \right)$,
${\cal O}\left( {MN_t^{RF}{N_r}{N_t}} \right)$ and
${\cal O}\left( {MN_t^{RF}{N_r}{N_t}} \right)$, respectively.
 Let ${I_{o}}$ and ${I_{i}}$ be the numbers of outer and inner iterations of Algorithm~\ref{alg3}, and the complexity of the algorithm can be expressed as:
$2{\cal O}\left( {{I_i}{I_o}MN_t^{RF}{N_r}{N_t}} \right)+{\cal O}\left( {{I_o}MN_t^{RF}{N_r}{N_t}} \right)+{\cal O}\left( {{I_o}MN_t^{RF}{N_s}{N_t}} \right)+{\cal O}\left( {{I_o}MN_t^{RF}{N_r^{RF}}{N_t}} \right)$. 

	For the MU-MISO scenario, the WMMSE-PMO algorithm proposed for the general MU-MIMO case remains applicable. However, the FP-PMO algorithm specifically designed for the MU-MISO case offers lower computational complexity compared to the WMMSE-PMO algorithm. The proof is provided as follows:
	As we can see, the complexity of FP-PMO can be expressed as:
	2${\cal O}\left( {{I_{in}}{I_{out}}{M^2}{N_r}{N_t}} \right)+{\cal O}\left( {{I_{out}}{M^4}} \right)$. 
	And the complexity of the WMMSE-PMO algorithm can be expressed as:
	$2{\cal O}\left( {{I_i}{I_o}MN_t^{RF}{N_r}{N_t}} \right)+{\cal O}\left( {{I_o}MN_t^{RF}{N_r}{N_t}} \right)+{\cal O}\left( {{I_o}MN_t^{RF}{N_s}{N_t}} \right)+{\cal O}\left( {{I_o}MN_t^{RF}{N_r^{RF}}{N_t}} \right)$, which is higher than that of FP-PMO algorithm. This is because the WMMSE algorithm introduces additional auxiliary variables ${\bf{\Lambda }}$, which lead to increased computational complexity. 

\begin{algorithm}[t]
	\caption{ WMMSE-PMO Algorithm for Multi-user MIMO Systems }\label{alg3}
	\renewcommand{\algorithmicrequire}{\textbf{Input:}}
	\renewcommand{\algorithmicensure}{\textbf{Output:}}
	\begin{algorithmic}[1]
		\REQUIRE  ${\bf{H}}$ , $\sigma _i^2$, for $i = 1,2, \cdots M$. 
		\STATE Initialize ${{\bf{U}}_{\left[ 0 \right]}},{\bf{W}}_{RF\left[ 0 \right]},{{\bf{V}}_{\left[0 \right]}},{{\bf{W}}_{\left[0 \right]}},{{\bf{\Lambda }}_{\left[0 \right]}}$, set $k=0$.
		\REPEAT
		\STATE Update ${{\bf{U}}_{\left[k \right]}}$ according to (\ref{eq60}) and (\ref{eqm13}). 
		\STATE Update ${{\bf{\Lambda }}_{\left[ k \right]}}$ according to (\ref{eq63}).
		\STATE Update ${{\bf{W}}_{D\left[ k \right]}}$ according to (\ref{eq65}).
		\STATE Calculate ${\bf{W}}_{RF\left[ k \right]}$ and ${{\bf{V}}_{\left[ k \right]}}$ with Algorithm~\ref{alg:PMO};
		
		\STATE $k=k+1$.
		\UNTIL{the objective function of (\ref{qa}) converges.}
		\ENSURE ${\bf{U}},{\bf{W}}_{RF}^{},{\bf{V}},{\bf{W}},{\bf{\Lambda }}$ 
	\end{algorithmic}
\end{algorithm}

\section{Simulation Results}\label{5}
This section presents simulation results to evaluate the performance of the proposed algorithms for both MU-MISO and MU-MIMO systems. The results are compared with those of existing  algorithms, as well as the optimal fully digital (FD-OPT) benchmark algorithm.
By assuming the same noise power $\sigma _i^2 = \sigma _n^2$ at all receivers, and all the results are obtained by averaging over 1000 channel realizations.


\subsection{MU-MISO Scenario}
In the MU-MISO system, we set the simulation parameters as follows. We generate
the channel matrix according to (\ref{eq9}). The number of effective channel paths is $L = 4$ and the spacing between the intelligent surface elements is half-wavelength. We assume the number of RF chains at the BS is ${N_{RF}} = 8$ and it connects ${N_t} = 64$ RIMSA elements. The BS serves $M=4$ UEs, and each connects to a RIMSA with ${N_r} = 4$.

In order to demonstrate the performance gain of our proposed algorithm, the benchmark algorithms are as follows.
\begin{itemize}
\item[$\bullet$] Optimal full digital scheme (FD-OPT): This refers to the all-digital connection where each RIMSA element is connected to a RF chain. In this case, the sum rate maximization problem is a convex one and FP could be used to obtain the global optimal solution, which provides an upper bound on the achievable sum rate. 
\item[$\bullet$] Only precoding at the transmitter side (AO-MO): This is the precoding algorithm that maximizes the sum rate in \cite{ref14}, which is proposed for a DMA BS.
\item[$\bullet$] Majorization-Minimization (MM) based method to maximize the sum rate (MM-Alt-Opt) as proposed in \cite{n28}. This is proposed for a hybrid MIMO transceiver system.
\end{itemize}
\subsubsection{Performance Versus SNR}
Fig. \ref{fig_4} shows the sum rate  vs. SNR.  The FD-OPT algorithm as an upper bound outperforms any of the other schemes, and the proposed FP-PMO algorithms outperform the MM-Alt-Opt and AO-MO algorithm. The superiority over \cite{ref14} stems from the use of waveguide-based transmission in DMA, which leads to a non-tunable transmission matrix. Additionally, this work assumes the deployment of RIMSA at both the transmitting and receiving ends, thereby offering enhanced array gain.
The proposed FP-PMO algorithm serves as an excellent candidate for joint transceiver design, achieving both better performance and
low complexity.

\subsubsection{Convergence Performance}
\begin{figure}[t]
\centering
\includegraphics[width=3.0in]{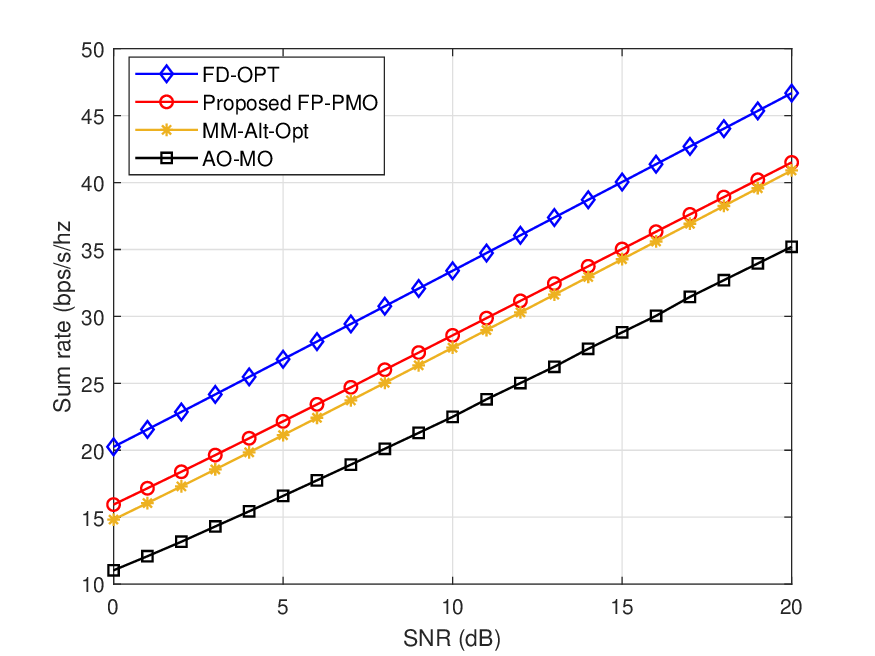}
\caption{Sum rate versus SNR under MU-MISO system(bit/s/Hz).}
\label{fig_4}
\end{figure}

\begin{figure}[t]
\centering
\includegraphics[width=3.0in]{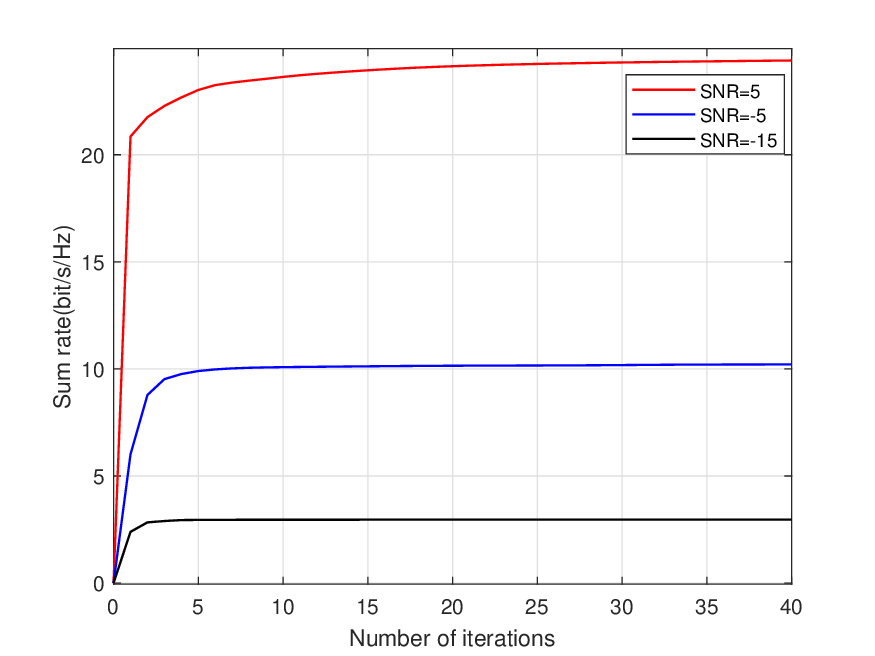}
\caption{Convergence performance of the proposed FP-PMO algorithm in different SNR (bit/s/Hz).}
\label{fig_5}
\end{figure}

Fig. \ref{fig_5} presents the convergence
behavior of our proposed FP-PMO algorithm. Simulation parameters are the same as those in Fig. \ref{fig_4}. We observe that the proposed algorithm converges after tens of iterations under different SNRs, which has  effective convergence performance in different environments.

\subsubsection{Performance Versus the Number of RIMSAs/RF Chains with Fixed RIMSA Aperture} 
\begin{figure}[t]
\centering
\includegraphics[width=3.0in]{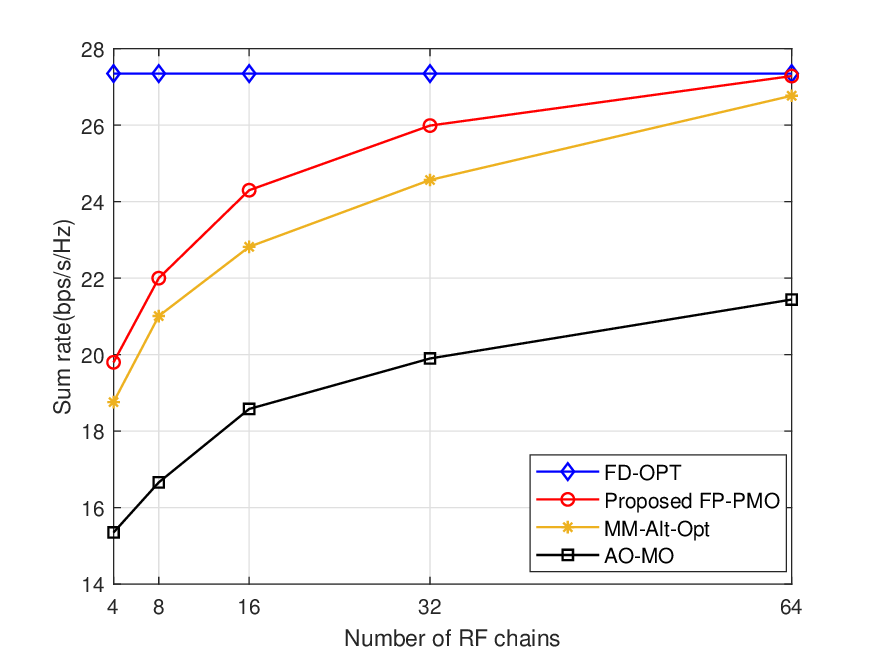}
\caption{Sum rate versus number of RF chains under MU-MISO system (bit/s/Hz).}
\label{fig_6}
\end{figure}

Fig. \ref{fig_6} shows the sum rate versus
the number of RF chains ${N_{RF}}$ when SNR = 5 dB. We observe that the FP-PMO algorithm outperform other algorithms. Furthermore,
 as ${N_{RF}}$ increases, the performance gap between the FP-PMO algorithm and FD-OPT becomes smaller. When the number of RF chains ${N_{RF}}$ is equal to ${N_t}$, the performance of the proposed algorithm is essentially the same as that of the FD-OPT algorithm. This is due to the physical structure. 
 

\begin{figure}[t]
\centering
\includegraphics[width=3.0in]{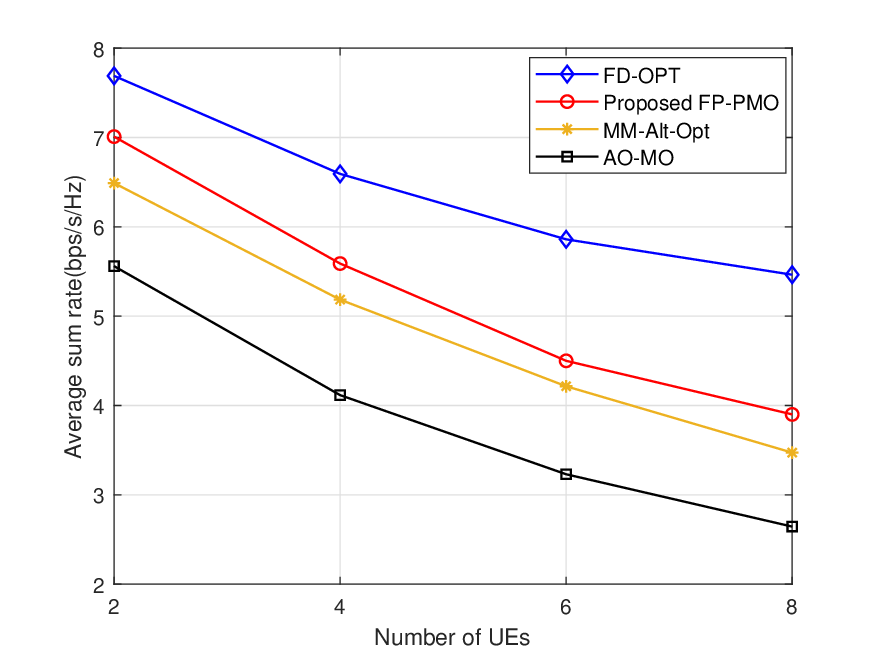}
\caption{{Average sum rate versus number of UEs under MU-MISO system (bit/s/Hz).}}
\label{fig_7}
\end{figure}

\subsubsection{Performance Versus the Number of UEs}
We explore the impact of the number of users served by the BS on the performance of our proposed algorithm with SNR = 5dB. We set $M = 2{\rm{,}}4{\rm{,}}6{\rm{,}}8$ and  other simulation conditions are the same as those in Fig. \ref{fig_4}. Fig. \ref{fig_7} shows that our proposed algorithm still outperforms the other algorithms. The above results demonstrate the robustness as well as the effectiveness of our proposed algorithm in MU-MISO scenarios.
\subsubsection{Imperfect CSI Analysis}
	To investigate the robustness of the proposed beamforming algorithm in practical scenarios, we consider a more realistic system model where the transmitter obtains only imperfect CSI. Specifically, the estimated CSI $\widehat {\bf{H}}$ is modeled as the sum of the true CSI ${\bf{H}}$ and a random estimation error ${\bf{E}}$:
	\begin{equation}
		\widehat {\bf{H}} = {\bf{H}} + {\bf{E}},
	\end{equation}
	where ${\bf{E}} \sim {\cal C}{\cal N}\left( {{\bf{0}},\sigma _e^2{{\bf{I}}_{{N_s}}}} \right)$ represents the channel estimation error, assumed to be independent and identically distributed (i.i.d.) complex Gaussian noise with variance $\sigma _e$. The parameter $\sigma _e$
	characterizes the quality of the CSI acquisition and can be adjusted to reflect different levels of imperfection.
	
\begin{figure}[t]
		\centering
		\includegraphics[width=3.0in]{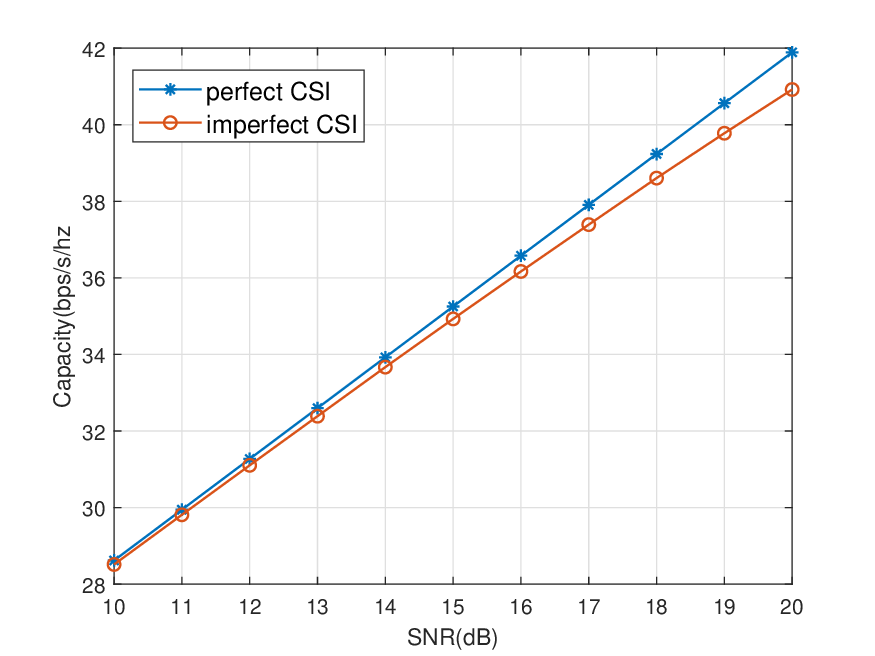}
		\caption{Performance of the algorithm  in the case of perfect CSI and imperfect CSI.}
		\label{fig_1}
	\end{figure}

	First, we fix the CSI error and simulate the effect of CSI error on the algorithm under different SNR. Fig. \ref{fig_1} illustrates the performance of the algorithm  in the case of perfect CSI and imperfect CSI with $\sigma _e=0.04$. It can be observed that the performance gap between the two scenarios is relatively small, which demonstrates the robustness and effectiveness of the proposed algorithm under imperfect CSI conditions.

	To gain a more comprehensive evaluation of the impact of imperfect CSI on the proposed algorithm, we conduct simulations by varying the error variance 
	$\sigma _e$
	from 0 (perfect CSI) to 0.1. The performance metric is the average achievable sum-rate over multiple Monte Carlo runs. The results are compared with the ideal CSI scenario to demonstrate the robustness of our design.
	
	\begin{table}
		[h!]
		\centering
		\caption{Performance Loss Varies with CSI Error}
		\resizebox{\linewidth}{!}{
			\begin{tabular}{|c|c|c|c|c|c|c|}
				\hline
				CSI error &0 & 0.02& 0.04& 0.06& 0.08& 0.1  \\\hline
				Performance Loss &0 & 0.13\% & 0.65\% & 1.25\%& 2.42\%& 3.9\%\\\hline

		\end{tabular}}
	\end{table}
	
	Table II indicate that although performance degradation occurs with increasing CSI error, the proposed algorithm maintains superior performance under moderate estimation errors, demonstrating a degree of robustness against CSI imperfection. This highlights the potential applicability of the proposed method in practical deployment scenarios where ideal CSI is unavailable.

\begin{figure}[!t]
\centering
\includegraphics[width=3.0in]{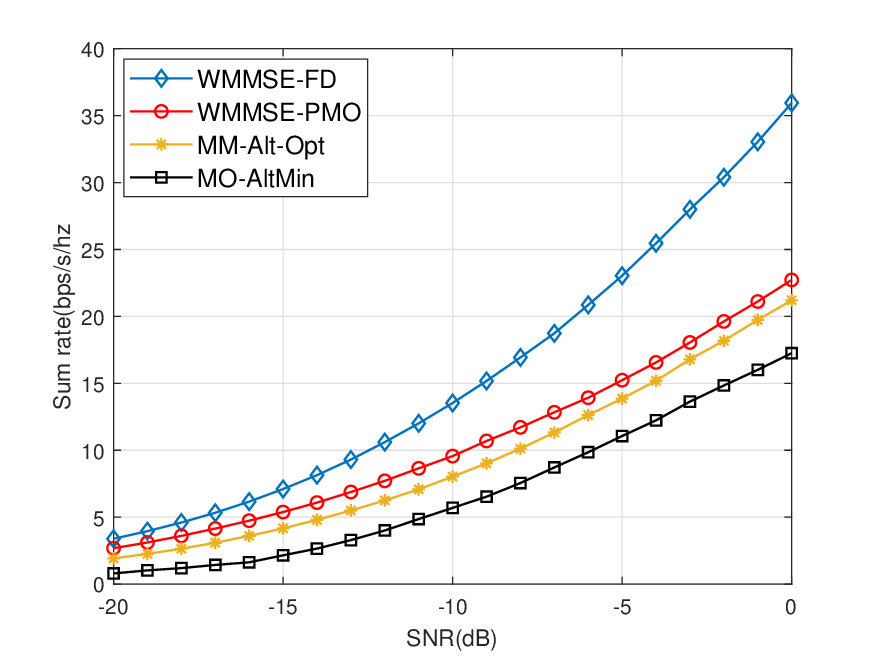}
\caption{Sum rate versus SNR under MU-MIMO system(bit/s/Hz).}
\label{fig_8}
\end{figure}
\begin{figure}[t]
	\centering
	\includegraphics[width=3.0in]{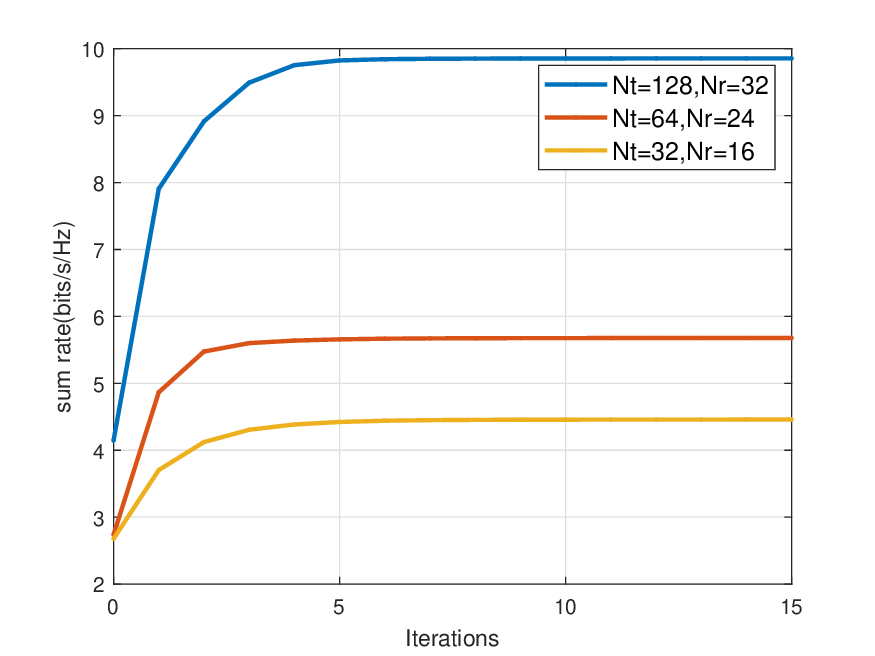}
	\caption{Convergence performance of the proposed WMMSE-PMO algorithm in different antennas..}
	\label{fig_9}
\end{figure}
\begin{figure}[!t]
\centering
\includegraphics[width=3.0in]{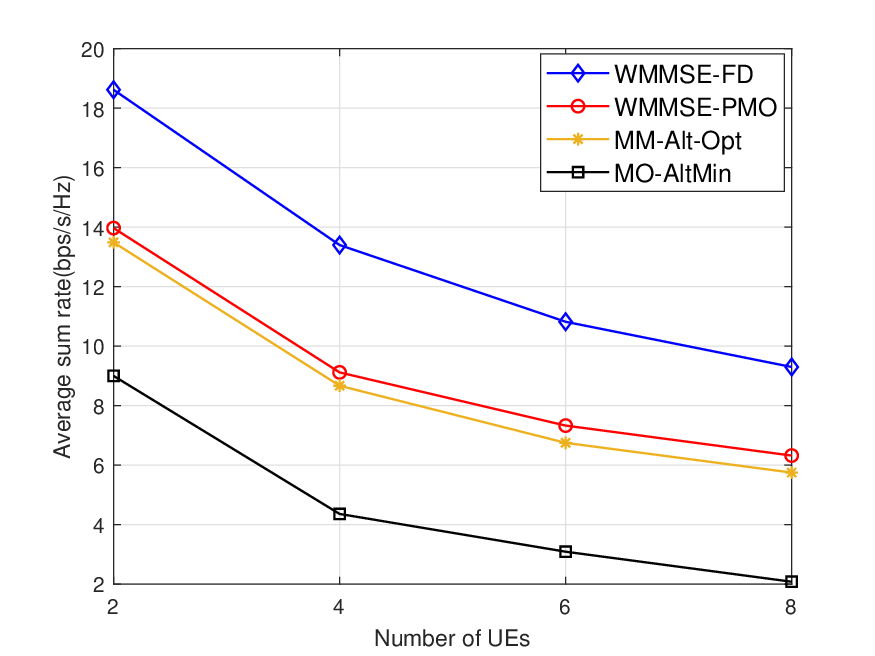}
\caption{{Average sum rate versus number of UEs under MU-MIMO system (bit/s/Hz).}}
\label{fig_11}
\end{figure}

\subsection{MU-MIMO Scenario}
For the MU-MIMO system,  We assume the number of RF chains at the BS is ${N_t^{RF}} = 16$ and it connects ${N_t} = 32$ RIMSA elements. The BS serves $M=4$ UEs, and each connected to an RIMSA with ${N_r} = 16$ and has ${N_r^{RF}} = 4$ RF chains.
We generate the channel matrix according to (\ref{eq9}). The number of effective channel paths is $L = 5$. The benchmark algorithms are as follows.
\begin{itemize}
\item[$\bullet$] Optimal full digital precoding (WMMSE-FD): This refers to the all-digital connection where each RIMSA element
is connected to a RF chain. In this case, the sum rate maximization problem can be solved by an AO algorithm where each subproblem is a convex one. This provides an upper bound on the achievable sum rate.
\item[$\bullet$] Traditional MO-AltMin algorithm: This
algorithm was proposed in  \cite{ref27}, where the optimal solution of full digital scenario is first obtained and then is approximated  by employing matrix factorization. 
\item[$\bullet$] MM-Alt-Opt method to maximize the sum rate as in \cite{n28}.
\end{itemize}
\subsubsection{Performance Versus SNR}
Fig. \ref{fig_8} shows the sum rate increases monotonically as
SNR increases. The WMMSE-FD scheme as an upper bound outperforms any of the other algorithms, and the performance of the proposed WMMSE-PMO algorithm is significantly better than the MO-AltMin and MM-Alt-Opt algorithm.

\subsubsection{Convergence Performance}
Fig. \ref{fig_9} presents the convergence behavior of our proposed WMMSE-PMO algorithm. The three curves in Fig. \ref{fig_9} show the converging behavior for ${N_t}=32,{N_r}=16$, ${N_t}=64,{N_r} =24$, and ${N_t}=128,{N_r}=32$, respectively under $SNR = -15$dB. Other simulation conditions are the same as those in Fig. \ref{fig_8}. We observe that the proposed algorithm converges after a few number of iterations for different number of antennas. The simulation results show that the our algorithm is effective.

\subsubsection{Performance Versus the Number of UEs}
Finally, we explore the impact of the number of users served by the BS on the performance of our proposed algorithm with SNR = 5dB. We set $M =2{\rm{,}}4{\rm{,}}6{\rm{,}}8{\rm{,}}$ and  other simulation parameters are the same as those in Fig. \ref{fig_8}. Fig. \ref{fig_11} shows that the average rate per UE in the system decreases as the number of UEs increases. This can be attributed to the increase in inter-user interference due to the increase in the number of UEs, as well as the degradation of the quality of service when the resource (e.g., BS power and number of antenna elements) keeps fixed. Our proposed algorithm still outperforms the other algorithm. The above results demonstrate the robustness as well as the effectiveness of our proposed algorithm in MU-MIMO scenarios.

 \begin{figure*}[t] 
\begin{equation}
\begin{aligned}
\label{eq81}
d{g_2} =& {\rm{Tr}}\left\{ {{{\left\{ {\left( {{\bf{\Phi }}\left( {{{\bf{\Phi }}^H} \odot {\bf{I}}_M^ - } \right)} \right) \odot {{\bf{I}}_M} + {{\bf{R}}_n}} \right\}}^{ - 1}}\left( {{{\bf{I}}_M} \odot \left( {\left( {d{{\bf{F}}_k}} \right){\bf{H}}{{\bf{V}}_k}{\bf{W}}\left( {{{\bf{\Phi }}^H} \odot {\bf{I}}_M^ - } \right)} \right)} \right)} \right\} \\
&{\rm{ - Tr}}\left\{ {{{\left( {\left( {{\bf{\Phi }}{{\bf{\Phi }}^H}} \right) \odot {{\bf{I}}_M} + {{\bf{R}}_n}} \right)}^{ - 1}}\left( {{{\bf{I}}_M} \odot \left( {d{{\bf{F}}_k}} \right){\bf{H}}{{\bf{V}}_k}{\bf{W}}{{\bf{\Phi }}^H}} \right)} \right\}\\
&\mathop {\rm{ = }}\limits^{\left( a \right)} {\rm{Tr}}\left\{ {{\bf{H}}{{\bf{V}}_k}{\bf{W}}\left( {{{\bf{\Phi }}^H} \odot {\bf{I}}_M^ - } \right)\left\{ {{{\left( {\left( {{\bf{\Phi }}\left( {{{\bf{\Phi }}^H} \odot {\bf{I}}_M^ - } \right)} \right) \odot {{\bf{I}}_M} + {{\bf{R}}_n}} \right)}^{ - 1}} \odot {{\bf{I}}_M}} \right\}\left( {d{{\bf{F}}_k}} \right)} \right\}\\
&{\rm{ - Tr}}\left\{ {{\bf{H}}{{\bf{V}}_k}{\bf{W}}{{\bf{\Phi }}^H}\left\{ {{{\left( {\left( {{\bf{\Phi }}{{\bf{\Phi }}^H}} \right) \odot {{\bf{I}}_M} + {{\bf{R}}_n}} \right)}^{ - 1}} \odot {{\bf{I}}_M}} \right\}\left( {d{{\bf{F}}_k}} \right)} \right\}{\rm{ = Tr}}\left\{ {{{\bf{A}}_k}\left( {d{{\bf{F}}_k}} \right)} \right\}.
\end{aligned}
\end{equation}
\begin{equation}
\begin{aligned}
\label{eq82}
d{g_2} =& {\rm{Tr}}\left\{ {{{\left\{ {\left( {{\bf{\Phi }}\left( {{{\bf{\Phi }}^H} \odot {\bf{I}}_M^ - } \right)} \right) \odot {{\bf{I}}_M} + {{\bf{R}}_n}} \right\}}^{ - 1}}\left( {{{\bf{I}}_M} \odot \left( {{{\bf{F}}_k}{\bf{H}}\left( {d{{\bf{V}}_k}} \right){\bf{W}}\left( {{{\bf{\Phi }}^H} \odot {\bf{I}}_M^ - } \right)} \right)} \right)} \right\} \\
&{\rm{ - Tr}}\left\{ {{{\left( {\left( {{\bf{\Phi }}{{\bf{\Phi }}^H}} \right) \odot {{\bf{I}}_M} + {{\bf{R}}_n}} \right)}^{ - 1}}\left( {{{\bf{I}}_M} \odot {{\bf{F}}_k}{\bf{H}}\left( {d{{\bf{V}}_k}} \right){\bf{W}}{{\bf{\Phi }}^H}} \right)} \right\}\\
&{\rm{ }}\mathop {\rm{ = }}\limits^{\left( b \right)} {\rm{Tr}}\left\{ {{\bf{W}}\left( {{{\bf{\Phi }}^H} \odot {\bf{I}}_M^ - } \right)\left\{ {{{\left( {\left( {{\bf{\Phi }}\left( {{{\bf{\Phi }}^H} \odot {\bf{I}}_M^ - } \right)} \right) \odot {{\bf{I}}_M} + {{\bf{R}}_n}} \right)}^{ - 1}} \odot {{\bf{I}}_M}} \right\}{{\bf{F}}_k}{\bf{H}}\left( {d{{\bf{V}}_k}} \right)} \right\}\\
&{\rm{  - Tr}}\left\{ {{\bf{W}}{{\bf{\Phi }}^H}\left\{ {{{\left( {\left( {{\bf{\Phi }}{{\bf{\Phi }}^H}} \right) \odot {{\bf{I}}_M} + {{\bf{R}}_n}} \right)}^{ - 1}} \odot {{\bf{I}}_M}} \right\}{{\bf{F}}_k}{\bf{H}}\left( {d{{\bf{V}}_k}} \right)} \right\}{\rm{ = Tr}}\left\{ {{{\bf{B}}_k}\left( {d{{\bf{V}}_k}} \right)} \right\}.
\end{aligned}
\end{equation}
 	\centering
 	{\noindent} \rule[-10pt]{18cm}{0.05em}
\end{figure*}

\section{Conclusion}\label{6}
This paper proposed an intelligent metasurface based RIMSA array and the corresponding transceiver scheme. With both RIMSA equipped at the BS and UEs,  downlink MU-MISO and MU-MIMO systems are investigated and optimized to maximize the sum rates. For the MU-MISO scenario,  we simplify the objective function by eliminating the summation term through matrix operations and propose an AO algorithm leveraging the FP and PMO techniques to solve this problem. For the MU-MIMO case, we extend the approach by introducing an AO algorithm based on PMO and WMMSE methods. Simulation results demonstrate that the proposed algorithm  outperforms existing baseline algorithms. Additionally, the algorithm exhibits fast convergence and robust performance across various scenarios, highlighting its effectiveness and practical viability.

	While this work primarily investigates the optimization of the sum rate in MU-MISO and MU-MIMO systems, future research can explore the deployment of RIMSA in more complex and dynamic communication scenarios.
	Moreover, further investigation is needed into the integration of RIMSA with practical signal processing modules such as channel estimation and direction-of-arrival (DoA) estimation, which are crucial for enabling adaptive and accurate beamforming in dynamic and uncertain environments. 
	Potential extensions include integrated air-space-ground networks, unmanned aerial vehicle (UAV)-assisted relay systems, and low Earth orbit (LEO) satellite communications. These environments impose stringent constraints on device size, power consumption, and real-time responsiveness, thereby underscoring the significance of RIMSA in enabling low-overhead, high-flexibility wireless communication architectures.
	In addition, the joint design of RIMSA and transceiver architectures may offer new opportunities for improving energy efficiency and enhancing physical layer security, especially in resource-constrained or hostile wireless environments.

\appendices
	 \section{ Calculate ${{\bf{A}}_k}$ and ${{\bf{B}}_k}$ }
Differentiating the function $g_2$ with ${{\bf{F}}_k}$ and ${{\bf{V}}_k}$, we can get  (\ref{eq81}) and  (\ref{eq82}) at the top of the next page, where $\left( a \right)$ and $\left( b \right)$ follow from the fact that ${\rm{Tr}}\left( {{{\bf{A}}^T}\left( {{\bf{B}} \odot {\bf{C}}} \right)} \right) = {\rm{Tr}}\left( {\left( {{{\bf{A}}^T} \odot {{\bf{B}}^T}} \right){\bf{C}}} \right)$ and ${\rm{Tr}}\left( {{\bf{ABC}}} \right) = {\rm{Tr}}\left( {{\bf{CAB}}} \right)$.
\section{ Proof of equivalence between (\ref{a}) and (\ref{qa}) }
 	First, because the objective function in (\ref{qa}) is convex with respect to ${{\bf{U}}_{{i}}}$ and  ${{{\bf{\Lambda }}_{{i}}}}$. Therefore, to minimize (\ref{qa}), the optimal ${{\bf{U}}_{{i}}}$ and  ${{{\bf{\Lambda }}_{{i}}}}$ can be obtained from (\ref{eq60}) and (\ref{eq63}), respectively. Substituting the  ${\bf{U}}_{{i}}^{mmse}$ and ${\bf{\Lambda }}_{{i}}^{opt}$ into (\ref{qa}), We can obtain the following equivalent optimization problem:   
 	\begin{equation}
 		\begin{aligned}
 			\label{eq52}
 			\mathop {\max }\limits_{{\bf{V}}{\rm{,  }}{\bf{W}}_{RF}^{\left( i \right)}{\rm{,}}{\bf{W}}_D\forall i} {\rm{   }}&\sum\limits_{i{\rm{ = 1}}}^M {\left( {\log \left| {{{\left( {{\bf{E}}_i^{mmse}} \right)}^{ - 1}}} \right|} \right)}     \\
 			\mbox{s.t.}\quad\quad
 			&{\rm{Tr}}({{\bf{W}}_D}{\bf{W}}_D^H) \le {P \mathord{\left/
 					{\vphantom {P N}} \right.
 					\kern-\nulldelimiterspace} K},{\rm{ }}\forall k   \\
&{{\bf{V}}} \in {{\cal V}_M},{\bf{W}}_{RF}^{\left( i \right)} \in {{\cal W}_M}.
 		\end{aligned}
 	\end{equation}
 	In the case where ${{\bf{U}}_{{i}}}$ is reversible, that is, when $N_r^{RF} = {N_s}$, we get
 	(\ref{eq53}), 
 	\begin{figure*}[ht] 
 		\begin{equation}
 			\begin{aligned}
 				\label{eq53}
 				\log \left| {{{\left( {{\bf{E}}_{{i}}^{mmse}} \right)}^{ - 1}}} \right|=& \log \det \left( {\frac{1}{{{{\bf{I}}_{{N_s}}} - {\bf{W}}_i^H{{\bf{V}}^H}{\bf{H}}_i^H{{\left({\bf{W}}_{RF}^{\left( i \right)} \right)}^H}{\bf{J}}_i^{ - 1}{\bf{W}}_{RF}^{\left( i \right)}{{\bf{H}}_i}{\bf{V}}{\bf{W}}_i}}} \right) \\
 				&\mathop {\rm{ = }}\limits^{(c)} {\rm{ - }}\log \det \left( {{{\bf{I}}_{{N_s}}} - {\bf{W}}_{RF}^{\left( i \right)}{{\bf{H}}_{{i}}}{{\bf{V}}}{{\bf{W}}_{{i}}}{\bf{W}}_{{i}}^H{\bf{V}}^H{\bf{H}}_{{i}}^H{{\left( {\bf{W}}_{RF}^{\left( i \right)} \right)}^H}{\bf{J}}_{{i}}^{ - 1}} \right)\\
 				&=\log \det \left( {{{\bf{I}}_{{N_s}}} + \frac{{{\bf{W}}_{RF}^{\left( {{i}} \right)}{{\bf{H}}_{{i}}}{{\bf{V}}}{{\bf{W}}_{{i}}}{\bf{W}}_{{i}}^H{\bf{V}}^H{\bf{H}}_{{i},}^H{{\left( {{\bf{W}}_{RF}^{\left( {{i}} \right)}} \right)}^H}}}
 					{\sum\limits_{\scriptstyle j = 1\hfill\atop
 							\scriptstyle j \ne i\hfill}^M {{\bf{W}}_{RF}^{\left( i \right)}{{\bf{H}}_i}{\bf{V}}{{\bf{W}}_j}{\bf{W}}_j^H{{\bf{V}}^H}{\bf{H}}_i^H{{\left( {\bf{W}}_{RF}^{\left( i \right)}\right)}^H}}  + \sigma _n^2{\bf{W}}_{RF}^{\left( i \right)}{{\left( {\bf{W}}_{RF}^{\left( i \right)} \right)}^H}}} \right)
 				{\rm{  = }}{r_{{i}}}.
 			\end{aligned}
 		\end{equation}
 		\centering
 		{\noindent} \rule[-10pt]{18cm}{0.05em}
 	\end{figure*}
 	where $\left( c \right)$ follows from the fact that $\det \left( {{\bf{I}} + {\bf{AB}}} \right) = \det \left( {{\bf{I}} + {\bf{BA}}} \right)$. Substituting (\ref{eq53}) into (\ref{eq52}) completes the proof.




\begin{thebibliography}{99}
	\balance

		\bibitem{ref1}
	D. C. Nguyen \emph{et al.}, ``6G internet of things: A comprehensive survey," \emph{ IEEE Int. Things J.}, vol. 9, no. 1, pp. 359-383, Jan. 2022.
	
		\bibitem{ref2}
C. -X. Wang  \emph{et al.}, ``On the road to 6G: Visions, requirements, key technologies, and testbeds," \emph{ IEEE Commun. Surv. Tuts.}, vol. 25, no. 2, pp. 905-974,  Second Quart. 2023.
 
	
	\bibitem{ref3}
	M. Agiwal, A. Roy and N. Saxena, ``Next generation 5G wireless networks: A comprehensive survey," \emph {IEEE Commun. Surv. Tuts.}, vol. 18, no. 3, pp. 1617-1655, Third Quart. 2016.
	
	\bibitem{ref4}
	S. Hu, F. Rusek and O. Edfors, ``Beyond massive MIMO: The potential of positioning with large intelligent surfaces," \emph { IEEE Trans. Signal Process.}, vol. 66, no. 7, pp. 1761-1774, Apr. 2018.
	
		\bibitem{refn8}
	J. Mo, A. Alkhateeb, S. Abu-Surra and R. W. Heath, ``Hybrid architectures with few-bit ADC receivers: Achievable rates and energy-rate tradeoffs," \emph{ IEEE Trans. Wireless Commun.}, vol. 16, no. 4, pp. 2274-2287, Apr. 2017.
	
	\bibitem{ref6}
	N. Shlezinger, G. C. Alexandropoulos, M. F. Imani, Y. C. Eldar and D. R. Smith, ``Dynamic metasurface antennas for 6G extreme massive MIMO communications," \emph{ IEEE Wireless Commun.}, vol. 28, no. 2, pp. 106-113, Apr. 2021.
	
	\bibitem{ref7}
	R. Deng \emph{et al.}, ``Reconfigurable holographic surfaces for future wireless communications," \emph { IEEE Wireless Commun.}, vol. 28, no. 6, pp. 126-131, Dec. 2021.
	
	
	\bibitem{ref8}
	T. Gong \emph{et al.},``Holographic MIMO communications: Theoretical foundations, enabling technologies, and future directions," \emph{ IEEE Commun. Surv. Tuts.}, vol. 26, no. 1, pp. 196-257, First Quart. 2024.
	
	\bibitem{ref9}
	B. Di, H. Zhang, L. Song, Y. Li, Z. Han and H. V. Poor, ``Hybrid beamforming for reconfigurable intelligent surface based multi-user communications: Achievable rates with limited discrete phase shifts,"  \emph{IEEE J. Sel. Areas Commun.}, vol. 38, no. 8, pp. 1809-1822, Aug. 2020.
	
	\bibitem{ref12}
	Q. Wu and R. Zhang, ``Towards smart and reconfigurable environment: Intelligent reflecting surface aided wireless network," \emph{ IEEE Commun. Mag.}, vol. 58, no. 1, pp. 106-112, Jan. 2020.
	
	\bibitem{refris}
	Y. Liu \emph{et al.}, ``Reconfigurable intelligent surfaces: Principles and opportunities," \emph{ IEEE Commun. Surv. Tuts.}, vol. 23, no. 3, pp. 1546-1577, Third Quart. 2021.
	
	\bibitem{refris1}
	S. Zeng \emph{et al.}, ``Reconfigurable intelligent surfaces in 6G: Reflective, transmissive, or both?," \emph{ IEEE Commun. Lett.}, vol. 25, no. 6, pp. 2063-2067, Jun. 2021.
	
	\bibitem{ref19}
	Q. Wu and R. Zhang, ``Intelligent reflecting surface enhanced wireless network via joint active and passive beamforming," \emph{IEEE Trans. Wireless Commun.}, vol. 18, no. 11, pp. 5394-5409, Nov. 2019.
	
	\bibitem{ref20}
	Q. Wu and R. Zhang, ``Joint active and passive beamforming optimization for intelligent reflecting surface assisted SWIPT under QoS constraints," \emph{IEEE J. Sel. Areas Communi.}, vol. 38, no. 8, pp. 1735-1748, Aug. 2020.
	
	\bibitem{ref21}
	S. Zhang and R. Zhang, ``Capacity characterization for intelligent reflecting surface aided MIMO communication," \emph{ IEEE J. Sel. Areas Commun.}, vol. 38, no. 8, pp. 1823-1838, Aug. 2020.
	
	
	\bibitem{ref22}
	C. Huang, A. Zappone, G. C. Alexandropoulos, M. Debbah and C. Yuen, ``Reconfigurable intelligent surfaces for energy efficiency in wireless communication," \emph{ IEEE Trans. Wireless Commun.}, vol. 18, no. 8, pp. 4157-4170, Aug. 2019.
	
	\bibitem{ref24}
	L. Dong and H.-M. Wang, ``Enhancing secure MIMO transmission via
	intelligent reflecting surface,” \emph{IEEE Trans. Wireless Commun.}, vol. 19,
	no. 11, pp. 7543–7556, Nov. 2020.

	
	\bibitem{ref25}
	L. Dong and H. -M. Wang, ``Enhancing secure MIMO transmission via intelligent reflecting surface," \emph{IEEE Trans. Wireless Commun.}, vol. 19, no. 11, pp. 7543-7556, Nov. 2020.
	
	\bibitem{refbai}
	J. Bai, H. -M. Wang and P. Liu, ``Robust IRS-aided secrecy transmission with location optimization," \emph{IEEE Trans. Commun.}, vol. 70, no. 9, pp. 6149-6163, Sept. 2022.

	
	\bibitem{reflv}
	W. Lv, J. Bai, Q. Yan and H. M. Wang, ``RIS-assisted green secure communications: Active RIS or passive RIS?," \emph{ IEEE Wireless Commun. Lett.}, vol. 12, no. 2, pp. 237-241, Feb. 2023.
	
		
	\bibitem{refD}
	L. Dong, H. -M. Wang and J. Bai, ``Active reconfigurable intelligent surface aided secure transmission," \emph{ IEEE Trans. Veh. Techn.}, vol. 71, no. 2, pp. 2181-2186, Feb. 2022.
	
			\bibitem{Transris3}
	J. Y. Lau and S. V. Hum, ``Analysis and characterization of a multipole reconfigurable transmitarray element," \emph{ IEEE Trans. Antennas and Prop.}, vol. 59, no. 1, pp. 70-79, Jan. 2011.
	
	\bibitem{ref11}
N. Shlezinger, O. Dicker, Y. C. Eldar, I. Yoo, M. F. Imani and D. R. Smith, ``Dynamic metasurface antennas for uplink massive MIMO systems," \emph{ IEEE Trans. Commun.}, vol. 67, no. 10, pp. 6829-6843, Oct. 2019.
	
	{	\bibitem{FIM}
	J. An \emph {et al.}, ``Flexible intelligent metasurfaces for downlink multiuser MISO communications," \emph { IEEE Trans. Wireless Commun.}, vol. 24, no. 4, pp. 2940-2955, April 2025.
	
	\bibitem{SIM1}
	J. An, C. Yuen, C. Xu, H. Li, D. W. K. Ng, M. Di Renzo, M. Debbah,
	and L. Hanzo, ``Stacked intelligent metasurface-aided MIMO transceiver
	design," \emph{ IEEE Wireless Commun.}, vol. 31, no. 4, pp. 123–131, Aug. 2024.
	
	\bibitem{SIM2}
	J. An, C. Xu, D. W. K. Ng, G. C. Alexandropoulos, C. Huang, C. Yuen,
	and L. Hanzo, ``Stacked intelligent metasurfaces for efficient holographic
	MIMO communications in 6G," \emph{IEEE J. Sel. Areas Commun.},  vol. 41, no. 8, pp. 2380–2396, Aug. 2023.}
	
	\bibitem{nat1}
H. Tian, L. Xu, X. Li, W. Jiang and T. Cui, ``Integrated control of radiations and in-band co-polarized reflections by a single programmable metasurface," \emph{Advan. Funct. Materials}, vol. 33, p. 2302753, Sep. 2023.
	
	\bibitem{ref13}
	H. Wang \emph {et al.}, ``Dynamic metasurface antennas based downlink massive MIMO systems," \emph{2019 IEEE 20th International Workshop on Signal Processing Advances in Wireless Communications (SPAWC)}, Cannes, France, 2019.
	
	\bibitem{ref14}
	S. F. Kimaryo and K. Lee, ``Downlink beamforming for dynamic metasurface antennas," \emph{ IEEE Trans. Wireless Commun.}, vol. 22, no. 7, pp. 4745-4755, Jul. 2023.

	
	\bibitem{ref15}
J. Xu, L. You, G. C. Alexandropoulos, J. Wang, W. Wang and X. Gao, ``Dynamic metasurface antennas for energy efficient uplink massive MIMO communications," \emph{2021 IEEE Global Communications Conference (GLOBECOM)}, Madrid, Spain, 2021.
	
	\bibitem{ref16}
R. Deng, B. Di, H. Zhang, Y. Tan and L. Song, ``Reconfigurable holographic surface: Holographic beamforming for metasurface-aided wireless communications," \emph{IEEE Trans. Veh. Techn.}, vol. 70, no. 6, pp. 6255-6259, Jun. 2021.
	
	\bibitem{ref17}
	R. Deng, B. Di, H. Zhang, Y. Tan and L. Song, ``Reconfigurable holographic surface-enabled multi-user wireless communications: amplitude-controlled holographic beamforming," \emph{IEEE Trans. Wireless Commun.}, vol. 21, no. 8, pp. 6003-6017, Aug. 2022.
	
	\bibitem{ref18}
	R. Deng, B. Di, H. Zhang, H. V. Poor and L. Song, ``Holographic MIMO for LEO satellite communications aided by reconfigurable holographic surfaces," \emph{IEEE J. Sel. Areas Commun.}, vol. 40, no. 10, pp. 3071-3085, Oct. 2022.
	
	{\bibitem{RTPS3} A. Singh and M. K. Mandal, ``Electronically tunable reflection type phase shifters," \emph{IEEE Trans. Circ. Systems II: Express Briefs}, vol. 67, no. 3, pp. 425-429, March 2020.}
	
	{\bibitem{RTPS2}  C. Chang, J. Chen, C. Shen, M. Tsai and T. Tai, ``Reflection-type phase shifter integrated with tunable power attenuation mechanism for sub-6 GHz wireless applications," \emph{IEEE Access}, vol. 10, pp. 115532-115540, 2022.		
	
	
	\bibitem{RTPS1} 
	M. Jeong, M. Kang, K. Oh and O. Lee, ``Frequency reconfigurable dual-band reflection-type phase shifter with  $360^{\circ}$ phase shift range for 5G NR FR2 applications ," \emph{J. Integ. Circ.  Systems}, vol. 10, no. 4, Oct. 2024.}
	
	
		\bibitem{HYB1}
	K. B. Dsouza, K. N. R. S. V. Prasad and V. K. Bhargava, ``Hybrid precoding with partially connected structure for millimeter wave massive MIMO OFDM: A parallel framework and feasibility analysis," \emph{IEEE Trans. Wireless Commun.}, vol. 17, no. 12, pp. 8108-8122, Dec. 2018.
	
	\bibitem{HYB2}
	Y. Feng and Y. Jiang, ``Hybrid precoding for massive MIMO systems
	using partially-connected phase shifter network," \emph{2019 11th International Conference on Wireless Communications and Signal Processing
	(WCSP)}, pp. 1–6, Oct 2019.

	{\bibitem{sim}
Q. Li, M. El-Hajjar, C. Xu, J. An, C. Yuen, and
L. Hanzo, “Stacked intelligent metasurfaces for holographic MIMO-aided cell-free network," \emph{IEEE Trans.
	Commun.},  vol. 72, no. 11, pp. 7139–7151, 2024.
		
			\bibitem{sim1}
		Q. Li, M. El-Hajjar, Y. Sun, and L. Hanzo, ``Performance analysis of
		reconfigurable holographic surfaces in the near-field scenario of cell-free
		networks under hardware impairments," \emph{IEEE Trans. Wireless Commun.}, vol. 23, no. 9, pp. 11972–11984, Sep. 2024.}
			
	\bibitem{ref26}
	Q. Shi, M. Razaviyayn, Z. -Q. Luo and C. He, ``An iteratively weighted MMSE approach to distributed sum-utility maximization for a MIMO interfering broadcast channel," \emph{ IEEE Trans. Signal Process.}, vol. 59, no. 9, pp. 4331-4340, Sept. 2011.
	
	\bibitem{ref27}
	X. Yu, J. -C. Shen, J. Zhang and K. B. Letaief, ``Alternating minimization algorithms for hybrid precoding in millimeter wave MIMO systems,"\emph{ IEEE J. Sel. Topics Signal Process.}, vol. 10, no. 3, pp. 485-500, April. 2016.
	
	
	
	\bibitem{ref28}
	E. Björnson, M. Bengtsson and B. Ottersten, ``Optimal multiuser transmit beamforming: A difficult problem with a simple solution structure," \emph{IEEE Signal Process. Mag.}, vol. 31, no. 4, pp. 142-148, July. 2014.
	
	\bibitem{ref29}
	K. Shen and W. Yu, ``Fractional programming for communication systems—part II: Uplink scheduling via matching," \emph{IEEE Trans. Signal Process.}, vol. 66, no. 10, pp. 2631-2644,  May. 2018.
	
	\bibitem{ref30}
	L. Zhang, Y. Wang, W. Tao, Z. Jia, T. Song and C. Pan, ``Intelligent reflecting surface aided MIMO cognitive radio systems," \emph{ IEEE Trans. Veh. Techn.}, vol. 69, no. 10, pp. 11445-11457, Oct. 2020.
	
	\bibitem{ref31}
	P. Peter, “Riemannian geometry,” \emph{IEEE Access}, 2006.
	
	\bibitem{ref32}
	P.-A. Absil, M. Robert and S. Rodolphe, ``Optimization algorithms on matrix manifolds,” \emph{IEEE Trans. Wireless Commun.}, 2008.
	
	
	
		\bibitem{n28}
   S. Gong, C. Xing, Vincent K. N. Lau, S. Chen, and L. Hanzo, ``Majorization-minimization aided hybrid transceivers for MIMO interference channels," \emph{IEEE Trans. Signal Process.},
	vol. 68, pp. 4903–4918, Nov. 2020.

	
\end{thebibliography}
\end{document}